\documentclass[twocolumn]{aastex63_mod}

\usepackage{xspace}
\usepackage{amsmath}
\usepackage{hyperref}

\newcommand{\um}{\hbox{$\mu$m}\xspace}
\newcommand{\Ha}{H$\alpha$\xspace}
\newcommand{\Hb}{H$\beta$\xspace}
\newcommand{\OIII}{[\ion{O}{3}] $\lambda 5007$\xspace}
\newcommand{\OIIIsimp}{[\ion{O}{3}]\xspace}

\newcommand{\OII}{[\ion{O}{2}] $\lambda 3727$\xspace}

\newcommand{\prosp}{\texttt{Prospector}\xspace}

\usepackage{booktabs}
\usepackage{changepage}
\date{\today}
\shorttitle{Bayesian Population Model for Dust Attenuation}
\shortauthors{Nagaraj et al.}
\graphicspath{{./}{figures/}}
\begin{document}

\title{A Bayesian Population Model for the Observed Dust Attenuation in Galaxies}

\correspondingauthor{Gautam Nagaraj}
\email{gxn75@psu.edu}

\author[0000-0002-0905-342X]{Gautam Nagaraj}
\affil{Department of Astronomy \& Astrophysics, The Pennsylvania State University, University Park, PA 16802, USA}
\affil{Institute for Gravitation and the Cosmos, The Pennsylvania State University, University Park, PA 16802, USA}
\affiliation{Center for Computational Astrophysics, 162 Fifth Avenue, New York, NY, 10010, USA}

\author[0000-0002-1975-4449]{John C. Forbes}
\affiliation{Center for Computational Astrophysics, 162 Fifth Avenue, New York, NY, 10010, USA}

\author[0000-0001-6755-1315]{Joel Leja}
\affil{Department of Astronomy \& Astrophysics, The Pennsylvania State University, University Park, PA 16802, USA}
\affil{Institute for Computational \& Data Sciences, The Pennsylvania State University, University Park, PA, USA}
\affil{Institute for Gravitation and the Cosmos, The Pennsylvania State University, University Park, PA 16802, USA}

\author[0000-0002-9328-5652]{Daniel Foreman-Mackey}
\affiliation{Center for Computational Astrophysics, 162 Fifth Avenue, New York, NY, 10010, USA}

\author[0000-0003-4073-3236]{Christopher C. Hayward}
\affiliation{Center for Computational Astrophysics, 162 Fifth Avenue, New York, NY, 10010, USA}

\begin{abstract}

Dust plays a pivotal role in determining the observed spectral energy distribution (SED) of galaxies. Yet our understanding of dust attenuation is limited and our observations suffer from the dust-metallicity-age degeneracy in SED fitting (single galaxies), large individual variances (ensemble measurements), and the difficulty in properly dealing with uncertainties (statistical considerations). In this study, we create a population Bayesian model to rigorously account for correlated variables and non-Gaussian error distributions and demonstrate the improvement over a simple Bayesian model. We employ a flexible 5-D linear interpolation model for the parameters that control dust attenuation curves as a function of stellar mass, star formation rate (SFR), metallicity, redshift, and inclination. Our setup allows us to determine the complex relationships between dust attenuation and these galaxy properties simultaneously. Using \prosp fits of nearly 30,000 3D-HST galaxies, we find that the attenuation slope ($n$) flattens with increasing optical depth ($\tau$), though less so than in previous studies. $\tau$ increases strongly with SFR, though when $\log~{\rm SFR}\lesssim 0$, $\tau$ remains roughly constant over a wide range of stellar masses. Edge-on galaxies tend to have larger $\tau$ than face-on galaxies, but only for $\log~M_*\gtrsim 10$, reflecting the lack of triaxiality for low-mass galaxies. Redshift evolution of dust attenuation is strongest for low-mass, low-SFR galaxies, with higher optical depths but flatter curves at high redshift. Finally, $n$ has a complex relationship with stellar mass, highlighting the intricacies of the star-dust geometry. We have publicly released \href{https://github.com/Astropianist/DustE}{software} for users to access our population model.

% Yet SED fitting codes struggle to constrain dust attenuation due to degeneracies with metallicity and age. Ensemble measurements of dust attenuation are limited by the wide variety among individual galaxies, and modeling dust evolution, properties, and radiative transfer effects from first principles is extremely challenging. Furthermore, most dust attenuation studies use simple strategies like binning to study trends as a function of galaxy properties, often ignoring uncertainties.

\end{abstract}


\section{Introduction} \label{sec:intro}

One of the least understood components of baryonic matter in galaxies is dust. While constituting only a negligible fraction of the total mass \citep[e.g.,][]{Driver2018}, dust is important in interstellar medium (ISM) evolution and chemistry, including cooling and catalysis of molecule formation. It plays an important role in star and planet formation and reprocesses up to 30\% of the bolometric luminosity of a galaxy \citep[e.g.,][]{Draine2003rev,Draine2003}. 

Perhaps the most influential challenge dust presents to astrophysical studies is the significant amount of absorption and scattering of light it causes, especially in the ultraviolet (UV)---in a galaxy setting we call this dust attenuation. An estimate of these effects is needed in order to measure several crucial physical parameters of galaxies, including star formation rates (SFRs) In fact, studies of galaxies at high redshift find the effects of dust to be one of the most important sources of uncertainty \citep[e.g.,][]{Bouwens2012,Finkelstein2012,Oesch2013}. 

Many observational studies at high redshift have employed the ``Calzetti Law'' \citep{Calzetti2000} as a model for the dust attenuation curve (see \S \ref{sec:methods} for description of attenuation laws) as a function of wavelength \citep[e.g.,][]{Erb2006,Daddi2007,Mannucci2010,PengY2010,Elbaz2011,MadauDickinson2014} and/or used empirical relations between observed and intrinsic UV slope from \cite{Calzetti1997,Meurer1999,Calzetti2001} to infer dust attenuation from UV spectra \citep[e.g.,][]{Shapley2001,Bouwens2015}. The aforementioned flagship studies provide simple methods to approximate dust attenuation based on ample observations in the local universe, making them quite useful.

However, the non-universality of dust attenuation curves has been known for decades. While \cite{Stecher1965} found evidence for a 2175 \AA~feature in extinction curves of the Milky Way and Large Magellanic Cloud (a feature not found in the Calzetti Law), \cite{Prevot1984} found a steeper extinction curve with no evidence for the bump in the Small Magellanic Cloud. \cite{WittGordon2000} showed through radiative transfer modeling that clumpier dust distributions lead to flatter attenuation curves. Observational studies have found that galaxies exhibit a wide variety of dust attenuation curves \citep[e.g.,][]{Burgarella2005,Noll2009,Wild2011,Buat2012,Arnouts2013,Kriek2013,Reddy2015,Salim2016,Leja2017,Salim2018}. Furthermore, the details of dust attenuation for a given galaxy have been shown to greatly influence the measurement of other properties, such as SFRs and histories \citep[e.g.,][]{Kriek2013,Reddy2015,Shivaei2015,Salim2016,SalimNarayanan2020}.

Several studies have looked into ensemble trends in attenuation curves, both in the local universe \citep[e.g.,][]{Burgarella2005,Salim2016,Leja2017,Salim2018} and at high redshift \citep[e.g.,][]{Arnouts2013,Kriek2013,Salmon2016,Tress2018}. These studies have greatly improved our empirical understanding of dust attenuation, but there is still progress to be made on understanding the evolution of attenuation and the complex interplay among variables \citep[][and references therein]{SalimNarayanan2020}, which is something we begin to tackle in this work.

The measurement of dust attenuation curves from photometry and/or spectra is done through empirical or modeling approaches (see \citeauthor{SalimNarayanan2020} \citeyear{SalimNarayanan2020} for a detailed summary of the literature). Empirical methods often use templates \citep[e.g.,][]{Calzetti1994,Calzetti1997,Calzetti2000,Johnson2007,Reddy2015,Battisti2016,Battisti2017a}, in which galaxy spectra are binned by Balmer decrement, infrared excess, or other relevant quantities, and averaged in each bin to create templates. The template corresponding to little or no dust is used to calibrate attenuation curves. While template empirical methods are useful for learning about broad trends in galaxy populations, they struggle with the complex correlations between parameters that are present in galaxies.

% The template empirical methods are used to characterize populations of galaxies. However, theoretical models predict that individual galaxies will exhibit a wide variety of attenuation slopes and bump strengths, making ensemble trends and averages a fragile indicator of dust properties \citep{Narayanan2018}. Furthermore, galaxies with the same dust reddening often have different intrinsic spectra due to variations in their stellar and nebular properties. Therefore, while such measurements are useful, they cannot tease the complex relationships between properties at once.

This leads us to the modeling approach, which is dominated by spectral energy distribution (SED) analysis (see \citeauthor{Walcher2011} \citeyear{Walcher2011} and \citeauthor{Conroy2013} \citeyear{Conroy2013} for reviews on this subject). SED modeling integrates various simple stellar populations---collections of stars created simultaneously from (single) molecular clouds---over a star formation history (SFH) and computes the integrated SED. Of course, to match the observed SED, dust attenuation must be taken into account. The goal of the SED model is to find the dust attenuation (and possibly emission if the mid- and/or far-IR is included), SFH, and metallicity parameters that are consistent with the data. 

While SED fitting is powerful in its ability to extract physical parameters of a galaxy based on photometry and/or emission line strengths or spectra, it faces many challenges. Relevant to this project; differing ages, metallicities, and dust attenuation curves can produce similar effects in spectra, causing degeneracies \citep[e.g.,][]{Conroy2013,Santini2015}. 

% Accounting for active galactic nuclei (AGN) is also often difficult without diagnostics in the MIR or other telltale signs. Furthermore, stellar evolution models continue to have uncertainties in binary physics, blue stragglers, horizontal branch morphology, thermal pulse asymptotic giant branch, stellar rotation, etc.

% Furthermore, the underlying components of SED models themselves are far from perfect, with potentially large uncertainties in the thermally pulsing asymptotic giant branch \citep[e.g.,][]{Kriek2010} and inclusion (or not) of binary/multi-star interactions \citep[e.g.,][]{Eldridge2008}.

Bayesian techniques, especially parameter space exploration methods like Markov Chain Monte Carlo (MCMC) algorithms, can help highlight the covariances among parameters. Such approaches have become increasingly common in SED modeling efforts over the past two decades, including \texttt{CIGALE} \citep{Noll2009,Boquien2019}, \texttt{GALMC} \citep{Acquaviva2011}, \prosp \citep{Leja2017,Johnson2021}, \texttt{MCSED} \citep{Bowman2020}, and others. 

% With the application of a state-of-the-art SED modeling code on a rich data set, we can shed new light on dust attenuation as a function of various physical quantities like stellar mass, specific SFR (sSFR), metallicity, and axis ratio over cosmic time. 

\begin{figure*}
    \centering
    \resizebox{\hsize}{!}{
    \includegraphics{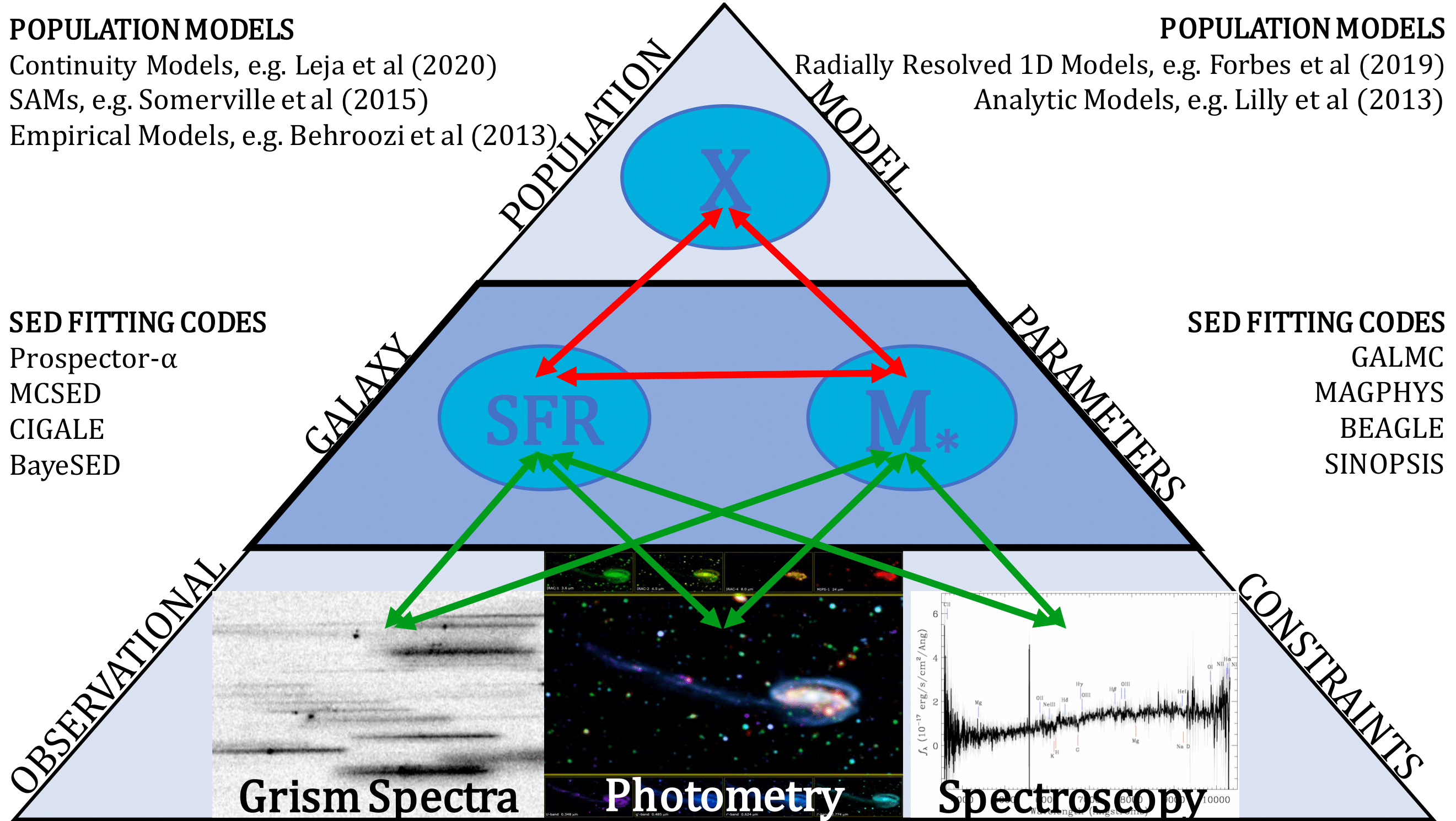}}
    \caption{Schematic showing the structure of a simple hierarchical model---the combination of all arrows to link the top and bottom layer---in the context of galaxy evolution. In this diagram, the ensemble property $X$ (e.g., a dust attenuation parameter) requires information about stellar mass and SFR of all of the individual galaxies (all arrows between the middle and top layers). Meanwhile, the information about stellar mass and SFR for individual galaxies is derived from the data, including spectroscopy and photometry, reflected by all arrows between the bottom and middle layers.}
    \label{fig:hiermodel}
\end{figure*}

While SED fitting is an effective way to derive properties of individual galaxies, the process of determining trends in ensembles of galaxies based on individual fits is often limited to binning properties by their maximum likelihood solutions and inspecting the results  \citep[e.g.,][]{Salim2018,Nagaraj2021a,Nagaraj2021b} due to the computational expenses of the alternative. In reality, observed SEDs are often noisy, and also, fitted parameters are highly correlated and subject to degeneracies, as mentioned earlier. Including proper treatment of correlated errors will lead to more accurate inferences and avoid potential methodological bias (see \citeauthor{ForemanMackey2014HierBay} \citeyear{ForemanMackey2014HierBay} for a detailed discussion). The method we adopt to account for covariances is population modeling, as portrayed in Figure \ref{fig:hiermodel}. 

The base layer of the pyramid represents observations. Technically, converting raw telescope data to photometry and spectroscopy would require another one or two pyramid layers, but for the scope of this paper we begin at the reduced data layer. In this work, the observations come from the 3D-HST photometric catalogs \citep{Skelton2014}. The middle layer represents measurements of physical properties of galaxies, which are directly input into our dust attenuation models (the top layer). In this work, we use \prosp parameter fits of 3D-HST galaxies. 

As a technical clarification, in this work, we use the arrows in the upward direction only, making our model strictly a population model. If we used the constraints on the population ``hyperparameters'' to inform better fits of individual galaxies in the sample, it would be a true hierarchical model.

% Our goal is to understand how populations of galaxies behave as a function of their physical properties, including stellar mass and SFR. However, these properties are typically measured indirectly, often through complex SED fitting algorithms, and are often correlated in complex ways.

% A hierarchical model properly takes into account the covariances between parameters for individual galaxies.

Only a few studies have conducted hierarchical studies of dust involving galaxy SEDs \citep{Kelly2012,Galliano2018}. These focus on just dust emission. In our study, we present the first population Bayesian modeling of dust attenuation. As described in \S \ref{sec:methods}, we model dust attenuation curves as a function of physical parameters of galaxies using posterior samples from \prosp fits of nearly 30,000 galaxies at a wide range of redshifts: $0.5<z<3.0$. As our galaxy sample is mass-complete, we do not need selection function modeling to make population-level inferences.

Through this effort, we highlight more complex trends in dust attenuation, pointing to gaps in our theoretical understanding. Our models, derived through statistically rigorous means, can be used by modelers who do not wish to incur the expense or systematic uncertainties of dust radiative transfer modeling. They can also be used to strengthen the prior for future SED studies of individual galaxies, allowing for tighter posterior constraints.

% SED models that use Bayesian MCMC techniques provide highest-likelihood solutions and posterior distributions for parameters. However, simply using either the highest-likelihood value or median value of each parameter to represent a given galaxy and using no error measurements or just a simple error cut (such as a $1-\sigma$ approximation) is not statistically justified and can lead to inaccurate inferences (see \citeauthor{ForemanMackey2014HierBay} \citeyear{ForemanMackey2014HierBay} for a detailed discussion on the correct use of statistics). Properly accounting for the covariances between parameters and individual galaxies requires a hierarchical model.

In \S \ref{sec:data} we describe the data we are using in our population model. This includes both observations (photometry and grism data) and measurements of physical parameters of galaxies using SED fitting. In \S \ref{sec:methods}, we describe the hierarchical Bayesian framework for our dust attenuation model and describe results from the tests used to demonstrate its accuracy. In \S \ref{sec:results}, we share our results when applying the population Bayesian model to the \prosp posterior samples and compare them to what we get when simply taking posterior medians for parameter values. We discuss some of the scientific statements we can make from the fitted models in \S \ref{sec:disc}. In \S \ref{sec:code}, we describe the public package we have created for users to adopt our models. Finally in \S \ref{sec:summary} we summarize our findings and suggest steps for the future.

\section{Data} \label{sec:data}

The 3D-HST Treasury program \citep{Brammer2012,Momcheva2016} used the G141 grism on the WFC3 camera on the \textit{Hubble Space Telescope} (\textit{HST}) to measure spectra of galaxies in 625 arcmin$^2$ of the CANDELS fields \citep{Grogin2011,Koekemoer2011}, a set of five highly studied patches of sky with a wealth of available observations. 

% With an average depth of 2 orbits, the program was able to reach limiting fluxes of $\sim 10^{-17}$ \flux. 

\cite{Skelton2014} used point spread function (PSF) matching techniques to properly combine photometry from 147 data sets at wavelengths $0.3-8.0$ \um for $\sim 200,000$ galaxies over a wide variety of redshifts. \cite{Momcheva2016} used state-of-the-art reduction techniques and grism-fitting software to calculate emission line fluxes and redshifts for nearly 80,000 galaxies. The most important lines included (at different redshift intervals) are \OII, \Hb, \OIII, and \Ha. For example, \Ha is in the grism wavelength range (1.08-1.67 \um) at redshifts $0.64<z<1.54$ whereas \OIII is present at $1.90<z<3.48$.

The data from the 3D-HST catalogs represent the bottom layer of the pyramid in Figure \ref{fig:hiermodel}. The actual inputs (middle layer) to our population model are the values of the physical parameters like stellar mass and SFR that are derived from the data catalogs via SED fitting. By including all posterior samples of the desired input variables from their SED fitting results, we are properly accounting for the uncertainties in the photometry as well as the correlations between SED-fitted parameters. The formalization of the process of creating a population model is presented in \S \ref{subsec:probmath}. 

% While the level of sophistication and implementation of novel techniques in SED modeling can greatly affect the accuracy of the results, another important factor is the rest-frame wavelength coverage of photometry and other data that go into each fit. For example, several 14-parameter models will be able to perfectly model an observed SED with five optical broadband photometric measurements even with small errors \citep[e.g., see discussion in][about fitting of photometry by various non-parametric models]{Leja2019a}. On the other hand, a large set of FUV-FIR photometry including narrow and/or medium bands in addition to various emission and/or absorption line fluxes will be much harder to reproduce. In other words, better and more extensive data helps break degeneracies in the physical parameters of the SED model.

For our project, we have used \prosp \citep{Leja2017,Johnson2021} fits of 3D-HST galaxies because the combination of ample broadband data and sophisticated SED modeling helps yield more accurate parameter fits. \prosp has undergone a number of posterior predictive checks to verify it produces accurate results, such as predicting the observed \Ha luminosities using constraints only from the photometry, \citep{Leja2017} and removing the discrepancy between the growth of the stellar mass density and the observed cosmic SFR density \citep{Leja2019}. By employing a non-parametric SFH, it allows more flexibility; indeed, the accumulation of more mass at earlier times, which would not be possible in standard parametric approaches, helped bridge the aforementioned disconnect \citep{Leja2019}. In addition, \prosp allows for custom or informative priors, permitting greater inclusion of physics and empirical results in the Bayesian framework. 

\cite{Leja2019,Leja2020} used \prosp to fit 63,427 objects from the 3D-HST catalogs at $0.5<z<3.0$. Of those, $\sim 30,000$ belong to a mass-complete regime in 3D-HST (based on the completeness estimation by \citeauthor{Tal2014} \citeyear{Tal2014}). In this work, we use the posterior samples from the mass-complete galaxies' fits. Restricting our analysis to the regime of completeness in 3D-HST allows us to make statements about a complete galaxy population without having to worry about selection effects. 

Specifically, the stellar mass, sSFR, metallicity, and dust attenuation measurements that we use in our modeling process are posterior samples from \prosp fits of the galaxies. Redshift measurements come from the 3D-HST team's analysis of grism spectra \citep{Momcheva2016}, though $\sim 70$\% of sources have only photometric redshifts available. These are not allowed to vary; in other words, each galaxy has a fixed redshift. Finally, axis ratios were derived using \texttt{GALFIT} \citep{Peng2002,Peng2010} and provided by \cite{vanderWel2014}. For these, posterior samples were generated using the maximum likelihood solutions as the means and $1-\sigma$ errors as Gaussian standard deviations.

% Metallicity and the dust attenuation parameters are fitted parameters of the model whereas stellar mass and sSFR are derived from fitted parameters.

The assumptions used in \prosp include the \cite{Chabrier2003} IMF and a flexible binned SFH. The dust attenuation law parameterization is detailed in \S \ref{subsec:daparam} (Equations \ref{eq:tau1}, \ref{eq:diffdust}, \ref{eq:drude}, \ref{eq:kc13}). Table 1 in \cite{Leja2019} lists the priors on all free parameters of their SED models.

% a \cite{Noll2009} dust attenuation law parameterization---a generalization of the \cite{Calzetti2000} law that includes a power-law component and a 2175 \AA~bump (see \S \ref{subsec:config} for full details)---with results from \cite{Kriek2013} dictating the relationship between the slope and bump strength. 

\section{Methodology} \label{sec:methods}

In this section we derive the equations governing our algorithms and explain how the models work. As there are many mathematical symbols defined throughout this section, we have created a glossary of terms (Table \ref{tab:glossary}) in Appendix \S \ref{sec:terms}. Before we dive into the details, we define dust attenuation laws below.

The reduction of light intensity through material along a given line of sight is called extinction. However, the overall effect dust has on light from an unresolved galaxy is called attenuation. Attenuation is a complex quantity that is a function of grain composition, dust-star-gas geometry, galaxy orientation, and the total amount of dust. 

One simple way to model dust attenuation is a screen model, in which the attenuation is modeled with a single dust screen extinguishing the light from the stars (and gas) of a galaxy. In this scenario, the complex effect of star-dust geometry is approximated by changes in the dust attenuation law, an equation for effective extinction vs wavelength. \cite{Calzetti1994,Calzetti2000} used such a method to find that dust attenuation curves of local starbursts have a relatively homogeneous form. As mentioned in \S \ref{sec:intro}, the Calzetti Law (\citeyear{Calzetti2000}) has played an integral role in dust attenuation studies over the last two decades.

As configured above, the screen model produces an effective dust attenuation curve. However, the success of the \cite{CharlotFall2000} two-component model suggests the utility of separating dust into diffuse dust and birth cloud dust. The two-component dust model is treated as a combination of two screens, one that affects all stars (diffuse) and one that affects only stars under 10 million years old (birth cloud). We use both effective and two-component dust screen models in our work. The details are provided in \S \ref{subsec:daparam}.

\subsection{Probability Theory for Population Model} \label{subsec:probmath}

We introduce the mathematical basis for the Bayesian model that we use to study dust attenuation as a function of various other properties, including stellar mass, SFR, galaxy inclination, metallicity, and redshift. Throughout this work, $p(\ldots)$ refers to the probability density of the quantity in the parentheses, possibly conditional on quantities behind a ``$|$''.

% Given the eclectic and complex landscape represented by the involved parameters, a thorough examination of probability theory is necessary to ensure we create a mathematically sound population Bayesian model. 

Our presentation largely follows that of  \cite{ForemanMackey2014HierBay}. We let $\{\mathbf{x}_k\}$ represent the entire set of data (mainly photometry) from the 3D-HST database used in the \prosp fits.  Next, $\{\mathbf{w}_k\}$ is the set of all the parameters of the galaxies (including stellar mass, SFR, $n$, etc.). More precisely, $\{\mathbf{w}_k\}$ represents the unknown ``true'' values of these parameters. In this notation, braces signify a set of galaxies, with the index ($k$) referring to individual galaxies. Finally, bold-face indicates a vector of parameters. 

In reality, not all vector components of $\{\mathbf{w}_k\}$ factor into every part of the computation process. For example, our population model does not consider Active Galactic Nuclei (AGN) fraction, though it is still a parameter in \prosp. To guarantee we are always considering the right subset of parameters, we can break down $\{\mathbf{w}_k\}$ into four components. 

\begin{itemize}
    \item $\{\mathbf{t}_k\}$ represents components like stellar mass that we include in our population Bayesian model that are not directly fitted in \prosp but are rather derived using other parameters.
    \item Next, we consider parameters that factor into our model and are directly fitted in \prosp (such as dust index $n$), which we will call $\{\mathbf{u}_k\}$.
    \item The last two components include parameters that are fitted in \prosp but are not directly considered in our model (such as the star formation history parameters). Those that go into calculating $\{\mathbf{t}_k\}$ we call $\{\mathbf{v}_k\}$.
    \item Finally, we christen the ``nuisance'' parameters in \prosp that do not have any bearing toward the population model $\{\mathbf{y}_k\}$. This includes parameters such as AGN fraction.
\end{itemize}

For the majority of this section, we will be focusing on probability distributions for individual galaxies. To keep the notation as uncluttered as possible, we will not include the subscript $k$, but it is still implied as the equations are valid for any given galaxy.

We must remember that $\mathbf{t}$ is a function of $\mathbf{u}$ and $\mathbf{v}$ since they are simply derived parameters. In other words, $\mathbf{t} = f(\mathbf{u},\mathbf{v})$, where $f$ is some deterministic function. Therefore, we can represent $\mathbf{w}$ as  $\left(\mathbf{u},\mathbf{v},\mathbf{y}\right)$. 

% In our case, $\{\mathbf{t}_k\}$ does not depend on $\{\mathbf{u}_k\}$, so we have simply

We let $\boldsymbol{\theta}$ be the parameters of our population model. Finally, we let $\boldsymbol{\alpha}$ represent the interim priors on $\mathbf{u},\mathbf{v},\mathbf{y}$ \prosp uses: $p(\mathbf{u},\mathbf{v},\mathbf{y}|\boldsymbol{\alpha})$. 

As mentioned in \S \ref{sec:data}, we are using the posterior samples from \prosp for nearly 30,000 galaxies to inform our population model. We can use the notation defined above to specify the distribution from which these samples are taken. 

\begin{equation} \label{eq:postdist}
    (\mathbf{u},\mathbf{v},\mathbf{y}) \sim p(\mathbf{u},\mathbf{v},\mathbf{y}|\mathbf{x},\boldsymbol{\alpha}) = \frac{p(\mathbf{x}|\mathbf{u},\mathbf{v},\mathbf{y})p(\mathbf{u},\mathbf{v},\mathbf{y}|\boldsymbol{\alpha})}{p(\mathbf{x}|\boldsymbol{\alpha})}
\end{equation}

% In this case, we are primarily interested in inferring the conditional distribution of the dust attenuation parameters as a function of the other quantities.

Our ultimate task is to derive the posterior distribution of $\boldsymbol{\theta}$, the parameters determining dust attenuation as a function of the physical properties for the full population of galaxies. In order to do that, we must determine a method to compute the probability density of the entire set of observations used in the \prosp analyses $\{\mathbf{x}_k\}$ given any $\boldsymbol{\theta}$, from which we can then apply Bayes' Theorem to compute the posterior density $p({\boldsymbol{\theta}}|\{\mathbf{x}_k\})$. 

Through the reasonable assumption that (the SEDs of) individual galaxies are independent of one another, we can write $p(\{\mathbf{x}_k\}|\boldsymbol{\theta}) = \prod_k p(\mathbf{x}_k|\boldsymbol{\theta})$ as long as the normalization is done properly. We begin with a statement of the law of total probability (once again omitting the subscript $k$).

\begin{equation} \label{eq:probexplicit}
    p(\mathbf{x}|\boldsymbol{\theta}) = \int p(\mathbf{x}|\mathbf{u},\mathbf{v},\mathbf{y})p(\mathbf{u},\mathbf{v},\mathbf{y}|\boldsymbol{\theta}) d\mathbf{u}d\mathbf{v}d\mathbf{y}
\end{equation}

% \begin{equation} \label{eq:totprob}
%     p(\{\mathbf{x}_k\}|\boldsymbol{\theta}) = \int p(\{\mathbf{x}_k\}|\{\mathbf{w}_k\},\boldsymbol{\theta})p(\{\mathbf{w}_k\}|\boldsymbol{\theta})d\{\mathbf{w}_k\}
% \end{equation}

In Equations \ref{eq:postdist} and \ref{eq:probexplicit}, we have made the reasonable assumption that the observations for a galaxy $\mathbf{x}$ do not depend directly on the interim prior parameters $\boldsymbol{\alpha}$ or the distribution of physical parameters $\boldsymbol{\theta}$, respectively, but rather just on the set of true values of the parameters $\mathbf{u},\mathbf{v},\mathbf{y}$. In fact, there may be shared systematic effects and modeling choices that could break the independence assumptions we have made, but we do not treat those effects here.

Equation \ref{eq:probexplicit} presents a high-dimensional integral that is difficult to compute as is, especially when considering that the calculation would need to be repeated for all galaxies in the sample. We will make approximations and adjustments to alleviate our task. First, we will multiply the integrand in Equation \ref{eq:probexplicit} by 1 in the following fashion.

\begin{multline} \label{eq:multby1}
    p(\mathbf{x}|\boldsymbol{\theta}) = \\ \int p(\mathbf{x}|\mathbf{u},\mathbf{v},\mathbf{y})p(\mathbf{u},\mathbf{v},\mathbf{y}|\boldsymbol{\theta}) \frac{p(\mathbf{u},\mathbf{v},\mathbf{y}|\mathbf{x},\boldsymbol{\alpha})}{p(\mathbf{u},\mathbf{v},\mathbf{y}|\mathbf{x},\boldsymbol{\alpha})} d\mathbf{u}d\mathbf{v}d\mathbf{y}
\end{multline}

\noindent We can use Equation \ref{eq:postdist} to simplify Equation \ref{eq:multby1}. We also use the fact that $p(\mathbf{x}|\boldsymbol{\alpha})$ is independent of the variables of integration.

\begin{equation} \label{eq:calc}
    \frac{p(\mathbf{x}|\boldsymbol{\theta})}{p(\mathbf{x}|\boldsymbol{\alpha})} = \int \frac{p(\mathbf{u},\mathbf{v},\mathbf{y}|\boldsymbol{\theta})}{p(\mathbf{u},\mathbf{v},\mathbf{y}|\boldsymbol{\alpha})} p(\mathbf{u},\mathbf{v},\mathbf{y}|\mathbf{x},\boldsymbol{\alpha}) d\mathbf{u}d\mathbf{v}d\mathbf{y}
\end{equation}

However, as we explained while defining terms, the population distribution we want to fit is actually $p(\mathbf{u},\mathbf{t}|\boldsymbol{\theta})$ rather than $p(\mathbf{u},\mathbf{v},\mathbf{y}|\boldsymbol{\theta})$. As defined earlier, $\mathbf{y}$ consists of variables such as AGN fraction that are not involved in our population model. In \prosp, these parameters have priors that are independent of the priors on $\mathbf{u}$ and $\mathbf{v}$. Furthermore, as $\mathbf{y}$ is unrelated to $\boldsymbol{\theta}$, the priors on our population model do not provide any additional constraints on it compared to the interim prior: $p(\mathbf{y}|\boldsymbol{\theta}) = p(\mathbf{y}|\boldsymbol{\alpha})$ Therefore, we have the following.

\begin{equation}
    \frac{p(\mathbf{u},\mathbf{v},\mathbf{y}|\boldsymbol{\theta})}{p(\mathbf{u},\mathbf{v},\mathbf{y}|\boldsymbol{\alpha})} = \frac{p(\mathbf{u},\mathbf{v}|\boldsymbol{\theta})}{p(\mathbf{u},\mathbf{v}|\boldsymbol{\alpha})} \frac{p(\mathbf{y}|\boldsymbol{\theta})}{p(\mathbf{y}|\boldsymbol{\alpha})} = \frac{p(\mathbf{u},\mathbf{v}|\boldsymbol{\theta})}{p(\mathbf{u},\mathbf{v}|\boldsymbol{\alpha})}
\end{equation}

\noindent In addition, we can write,

\begin{equation}
    \frac{p(\mathbf{u},\mathbf{v}|\boldsymbol{\theta})}{p(\mathbf{u},\mathbf{v}|\boldsymbol{\alpha})} = \frac{p(\mathbf{u},\mathbf{t}|\boldsymbol{\theta})}{p(\mathbf{u},\mathbf{t}|\boldsymbol{\alpha})}
\end{equation}

\noindent since the function $\mathbf{t}=f(\mathbf{u},\mathbf{v})$ does not depend on $\boldsymbol{\alpha}$ or $\boldsymbol{\theta}$, implying that the Jacobian of the transform in the numerator and denominator will be the same and therefore cancel in this ratio. From these considerations, we can rewrite Equation \ref{eq:calc} as follows.

\begin{equation} \label{eq:calcut}
    \frac{p(\mathbf{x}|\boldsymbol{\theta})}{p(\mathbf{x}|\boldsymbol{\alpha})} = \int \frac{p(\mathbf{u},\mathbf{t}|\boldsymbol{\theta})}{p(\mathbf{u},\mathbf{t}|\boldsymbol{\alpha})} p(\mathbf{u},\mathbf{v},\mathbf{y}|\mathbf{x},\boldsymbol{\alpha}) d\mathbf{u}d\mathbf{v}d\mathbf{y}
\end{equation}

We can numerically estimate $p(\mathbf{u},\mathbf{t}|\boldsymbol{\alpha})$ by sampling directly from \prosp with no data. We discuss this process in Appendix \ref{sec:prior} and show examples of interim prior probability distributions. Meanwhile, from Equation \ref{eq:postdist}, we know that the \prosp posterior samples are taken from the distribution $p(\mathbf{u},\mathbf{v},\mathbf{y}|\mathbf{x},\boldsymbol{\alpha})$. Therefore, we can use the Monte Carlo integral approximation to estimate the integral in Equation \ref{eq:calcut}. 

% that it provides an adequate estimate for the interim prior probability distribution

% Below, since $p(\mathbf{x}|\boldsymbol{\alpha}$ is a constant with respect to the variables of integration, we refer to it as $Z_{\boldsymbol{\alpha}}$.

\begin{equation}
    \frac{p(\mathbf{x}|\boldsymbol{\theta})}{p(\mathbf{x}|\boldsymbol{\alpha})} \approx \frac{1}{M} \sum_{m=1}^M \frac{\left( \mathbf{u}^{(m)},\mathbf{t}^{(m)} | \boldsymbol{\theta} \right)}{\left( \mathbf{u}^{(m)},\mathbf{t}^{(m)} | \boldsymbol{\alpha} \right)}
\end{equation}

\noindent To fit the set of all galaxies, based on the galaxy independence assumption stated earlier, we simply take the product of probabilities for all galaxies.

\begin{equation}
    \frac{p(\{\mathbf{x}_k\}|\boldsymbol{\theta})}{p(\{\mathbf{x}_k\}|\boldsymbol{\alpha})} = \prod_{k=1}^N \frac{p(\mathbf{x}|\boldsymbol{\theta})}{p(\mathbf{x}|\boldsymbol{\alpha})}
\end{equation}

\noindent The quantity $p(\{\mathbf{x}_k\}|\boldsymbol{\theta})$ is called the likelihood, which we will label $\mathcal{L}$. For computational purposes, we take the logarithm and ignore the constant $p(\{\mathbf{x}_k\}|\boldsymbol{\alpha})$ since it is independent of $\boldsymbol{\theta}$.

\begin{equation} \label{eq:compute_gen}
    \ln{\mathcal{L}} = \sum_{k=1}^N \ln \left( \sum_{m=1}^M \frac{p\left(\mathbf{t}_k^{(m)},\mathbf{u}_k^{(m)}|\boldsymbol{\theta}\right)}{p\left(\mathbf{t}_k^{(m)},\mathbf{u}_k^{(m)}|\boldsymbol{\alpha}\right)} \right)
\end{equation}

\noindent In \S \ref{subsec:moddetails}, we make Equation \ref{eq:compute_gen} more specific to our model.

\subsection{Dust Attenuation Parameterization} \label{subsec:daparam}

As mentioned in \ref{subsec:probmath}, the goal of this work is to learn how the dust attenuation parameters vary as a function of other galaxy properties, including stellar mass, SFR, stellar metallicity, axis ratio, and redshift. Before getting into the details of our population model, we will describe the parameterization of dust attenuation that we are using. \prosp uses a two-component model (modified from \citeauthor{CharlotFall2000} \citeyear{CharlotFall2000}): diffuse dust, affecting all sources in each galaxy, and birth cloud dust, affecting just the most recently formed stars. The birth cloud dust attenuation curve is modeled as a $\lambda^{-1}$ power law while the diffuse dust is modeled in the form introduced by \cite{Noll2009}. 

\begin{equation} \label{eq:tau1}
    \tau_{\lambda,1} = \tau_1 \left( \frac{\lambda}{5500 \AA} \right)^{-1} 
\end{equation}

\begin{equation} \label{eq:diffdust}
    \tau_{\lambda,2} = \frac{\tau_2}{4.05}\left(k'(\lambda)+D(\lambda) \right) \left( \frac{\lambda}{5500 \AA} \right)^{n} 
\end{equation}

\begin{equation} \label{eq:drude}
    D(\lambda) = \frac{E_b(\lambda \Delta \lambda)^2}{(\lambda^2-\lambda_0^2)^2 + (\lambda \Delta \lambda)^2}
\end{equation}

\begin{equation} \label{eq:kc13}
    \lambda_0,~\Delta \lambda,~E_b = 2175 \AA,~350 \AA,~0.85-1.9n 
\end{equation}

\noindent Here, $k'(\lambda)$ is the Calzetti Law \citep[see][]{Calzetti2000}. The formula $E_b=0.85-1.9n$ is given by \cite{Kriek2013}. The parameters that are allowed to vary are $\tau_1$, $\tau_2$, and $n$. However, \prosp puts a normal prior on $\tau_1/\tau_2$ with a mean $\mu=1$ and width $\sigma=0.3$ based on empirical evidence \citep[e.g.,][]{Calzetti1994,Price2014}. This prior places strong constraints on $\tau_1$. Therefore, we focus on $\tau_2$ (diffuse dust normalization) and $n$ (relating to the slope of the dust attenuation curve, with negative values indicating steeper than the Calzetti Law and positive values indicating shallower) as the primary dependent variables of the dust attenuation model.

In addition, we compute dust attenuation models for a one-component dust model with the same parameterization as diffuse dust (Equation \ref{eq:diffdust}). To ensure the same treatment in the population Bayesian modeling, we calculated the effective dust attenuation curve for every \prosp posterior sample by generating the dust-free and dust-included spectra using the \prosp parameter values and fitting an optical depth model with Equation \ref{eq:diffdust} using a basic curve fitting package from SciPy. The one-component model is useful in its simplicity, especially in cases where not much is known about the stellar populations comprising a given galaxy. 

% We show the direct comparison of diffuse dust to effective dust in Appendix \ref{sec:diffeff}.

\begin{figure*}
    \centering
    \resizebox{\hsize}{!}{
    \includegraphics{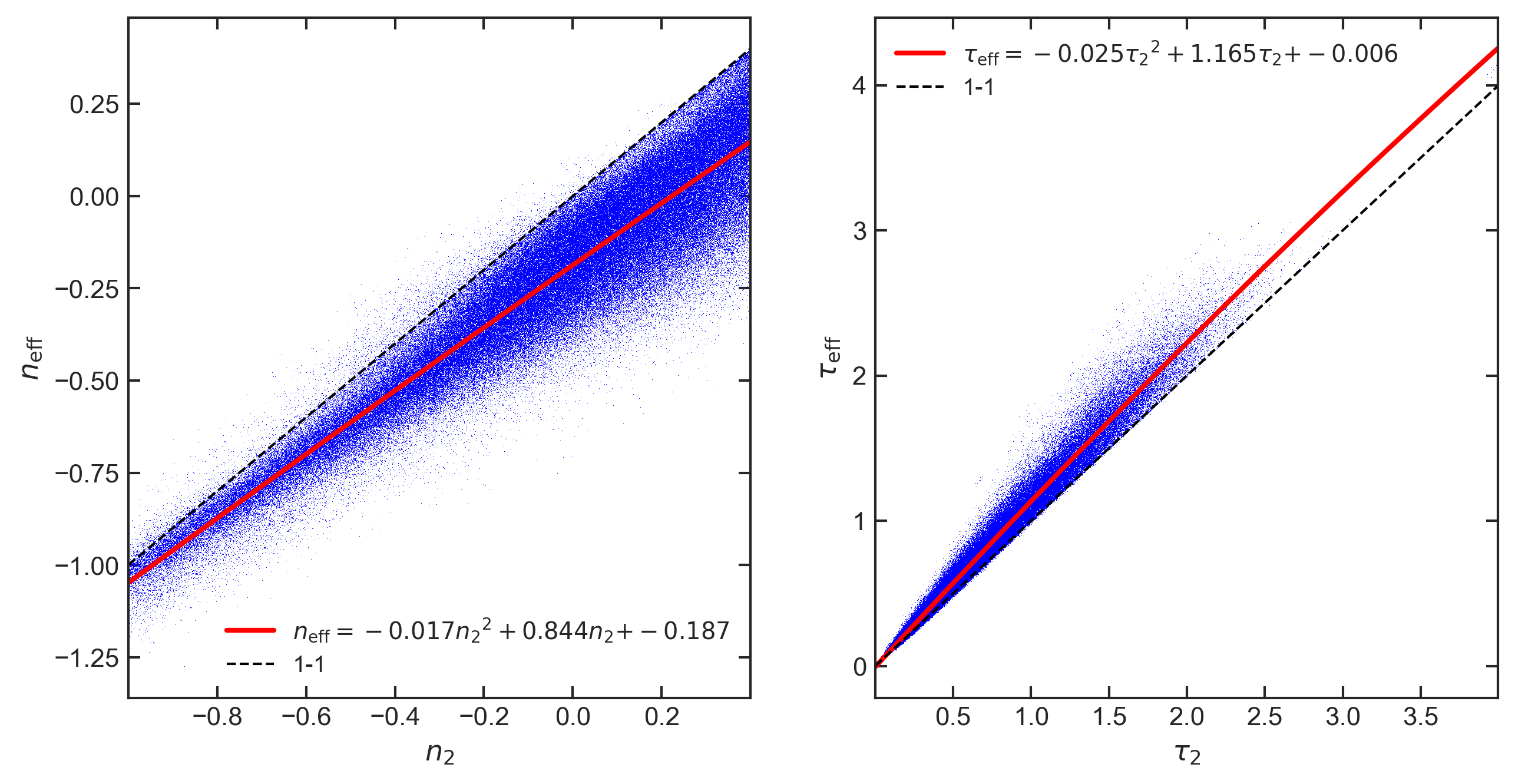}}
    \caption{Direct comparison of effective and diffuse dust attenuation curves. Left panel: slope parameter comparison. In agreement with \cite{Inoue2005}, \cite{Salim2018}, and \cite{SalimNarayanan2020}, we find that the effective slope is systematically steeper than the diffuse slope, which can be attributed to the unevenly distributed effects of birth cloud dust on UV light from a (star-forming) galaxy. Right panel: optical depth comparison. As expected, the effective optical depth is greater than the diffuse optical depth due to the contribution from birth cloud dust.}
    \label{fig:diffeff}
\end{figure*}

While we provide different models for single- and two-component dust models, it is useful and interesting to study the direct relationship between the two as we can learn more about how the stellar and nebular components contribute individually to the overall picture. In Figure \ref{fig:diffeff}, we provide the comparison between diffuse and effective dust. 

The left panel shows $n_{\rm eff}$ vs $n$ (diffuse). We see that the effective slope is systematically steeper than the diffuse dust slope, with a difference of $0.05-0.2$, which is consistent with the theoretical expectations and observational findings of \cite{Inoue2005},\cite{Salim2018}, and \cite{SalimNarayanan2020}, as stars affected by both diffuse and birth cloud dust will be heavily attenuated.

% As detailed in \S \ref{subsec:config}, in order to calculate the one-component dust attenuation parameters $n_{\rm eff}$ (slope) and $\tau_{\rm eff}$, we calculated the dust-free and dust-attenuated spectra for all \prosp posterior samples and fit for effective dust attenuation.

However, we do observe objects above the one-to-one line in Figure \ref{fig:diffeff}. As we did not impose the requirement that $n_{\rm eff} < n_{\rm diff}$, there are cases in which the fitted effective optical depth is considerably larger than the diffuse optical depth and the fitted effective slope is shallower than the diffuse slope. Nevertheless, such cases account for fewer than 1\% of all posterior samples, making them have minimal effect on the conclusions we draw throughout the paper.

% The explanation for this phenomenon is the following: in a (largely) star-forming population of galaxies, young stars dominate at short (UV) wavelengths and are affected by both diffuse and birth cloud dust. As a result, the UV output is drastically reduced compared to the optical, or in other words, a steeper attenuation law.

The fact that the best-fit slope for $n_{\rm eff}$ vs $n_{\rm diff}$ is not one is a bit more complex. As discussed later in \S \ref{subsec:AvSDisc}, higher optical depths are linked to shallower slopes \citep[][and references therein]{SalimNarayanan2020}. In addition, as we show in Figure \ref{fig:dust1} (\S \ref{sec:results}), birth cloud dust optical depth increases super-linearly with diffuse dust optical depth. Therefore, at lower optical depths, which are associated with steeper slopes, the difference between birth cloud dust and diffuse dust is smaller, leading to a smaller change between diffuse and effective slope.

The right panel shows $\tau_{\rm eff}$ vs $\tau_2$. The relationship between these two is straightforward: as long as the effective slope is not radically different than the diffuse slope, the effective dust optical depth must be greater than the diffuse dust optical depth. As optical depth increases, the birth cloud contribution increases, leading to a greater divergence between effective and diffuse.

\subsection{Interpolation Configuration} \label{subsec:config}

The functional form we have chosen for the dust attenuation curve is an N-D linear interpolation with an intrinsic Gaussian scatter. Given that the slope and normalization of the dust attenuation curve have been found to be correlated \citep[][and references therein]{SalimNarayanan2020}, we consider the joint modeling of the slope ($n$) and optical depth ($\tau$)\footnote{Henceforth, when generalizing for both one-component and two-component dust models, $n$ and $\tau$ without any subscripts will be used. For the one-component model specifically, we will use $n_{\rm eff}$ and $\tau_{\rm eff}$, and for two-component $n$ and $\tau_2$.} While the interpolation components of the model are considered independent, we treat the intrinsic scatter as a bivariate Gaussian with a possible covariance $\rho$ between $n$ and $\tau$. 

Below, we show the equations for our population model, where $I_1$ and $I_2$ are linear N-D interpolations, $\mathcal{Z}_1$ and $\mathcal{Z}_2$ are samples from the standard normal distribution, and $\sigma_1$ and $\sigma_2$ are the intrinsic widths of the relation in $n$ and $\tau_2$, respectively.

\begin{align}
    n &= I_1(\log~M_*, \log {\rm SFR}, ...) + \sigma_1 \mathcal{Z}_1 \label{eq:npoly} \\
    \tau &= I_2(\log~M_*, \log {\rm SFR}, ...) + \sigma_2 \left( \rho \mathcal{Z}_1 + \sqrt{1-\rho^2}\mathcal{Z}_2 \right) \label{eq:tau2poly}
\end{align}

In addition to this bivariate model, we compute univariate models in which $\tau$ is an independent variable and $n$ is the sole dependent variable (Equation \ref{eq:npoly}).

% Such models more explicitly require the correlation between $n$ and $\tau$. 

The parameters for the interpolation $I_1$ and $I_2$ are the values of $n$ and $\tau$, respectively, at a given set of grid points in the N-D independent parameter space. We employ unevenly spaced grids where points are closer together in regions where more data is present. This process helps maximize the utility of the data. We use five grid points per dimension, leading to 3125 total grid points for $n$ and $\tau$ (6250 parameters total) for the 5-D models.

\subsection{Modeling Details} \label{subsec:moddetails}

In \S \ref{subsec:probmath}, we laid out the mathematical foundation for our population Bayesian model. In \S \ref{subsec:config}, we described the model parameterization we adopted. In this section, we connect \S \ref{subsec:probmath} and \S \ref{subsec:config} and provide details about the modeling process itself. 

To run our population model, we use the Python package \texttt{pymc3} \citep{Pymc3}, which employs an efficient Hamiltonian Monte Carlo algorithm called No U-Turn Sampler \citep[NUTS;][]{HoffmanGelman2011} to move through the parameter space. We typically utilize four Markov chains, each with 1,000 tune steps (initial steps that are subsequently thrown away) and 1,000 accepted steps. In all cases for our simulated data (Appendix \ref{sec:simres}) and most cases for our actual results (\S \ref{sec:results}), these numbers are sufficient for chain convergence.

The model parameters are the values of $n$ and $\tau$ at the interpolation grid points (6250 parameters for the 5-D models), $\log \sigma_1$, $\log \sigma_2$, and $\rho$. The prior for $n$ and $\tau$ grid points is simply the \prosp prior for those variables: a uniform distribution between $[-1.0,0.4]$ and a truncated Gaussian distribution with $(\mu=0.3,~\sigma=1.0)$ and bounds $[0.0,4.0]$, respectively (with slightly different values for the one-component dust model). The priors for $\log \sigma_1$ and $\log \sigma_2$ are Student-T distributions with $(\nu=5,\mu=-3.0,\sigma=0.5)$ while the prior for $\rho$ is a uniform distribution in $[-1,1]$. For simplicity, we will refer to the $n$ and $\tau$ grid parameters using the interpolation functions $I_1$ and $I_2$.

If we let $I_1^*$, $I_2^*$, $\sigma_1^*$, $\sigma_2^*$, and $\rho^*$ represent the current model parameters, where $^*$ denotes the particular model, we can define the likelihood of the $m^{\rm th}$ posterior of the $k^{\rm th}$ galaxy in the sample in the following manner. Here, $I_{1,k,m}^* \equiv I_1^*(\log M_{*k,m},...)$, etc.

\begin{align}
    \mathcal{Z}_{k,m} &= \frac{{\left(n_{k,m}-I_{1,k,m}^*\right)}^2}{\sigma_1^2} + \frac{\left(\tau_{2,k,m}-I_{2,k,m}^*\right)^2}{\sigma_2^2} \nonumber \\ 
    & + \frac{\left(n_{k,m}-I_{1,k,m}^*\right) \left(\tau_{2,k,m}-I_{2,k,m}^*\right)}{\sigma_1 \sigma_2} \\
    \ln \mathcal{L}_{{\rm hyper},k,m} &= -0.5\frac{\mathcal{Z}_{k,m}}{1-\rho^2} - \ln{\left( \sigma_1 \sigma_2 \sqrt{1-\rho^2}\right)} \label{eq:like_hyper}
\end{align}

From Equation \ref{eq:compute_gen}, we see that to compute the true likelihood, we still need to account for the \prosp priors. As the process of computing the prior probabilities in a consistent manner with the likelihood in Equation \ref{eq:like_hyper} is not straightforward but is a minor concern compared to the details of the population model, we have kept details of our process in Appendix \ref{sec:prior}. 

Based on Equation \ref{eq:compute_gen}, we compute the contribution to the likelihood from the $k$th galaxy's $m$th \prosp sample %true likelihood as follows.

% Even though our models involved only up to five parameters from \prosp (with three of them being derived from the free parameters), the posterior samples themselves depend on all \prosp free parameters, as hinted by Equation \ref{eq:postsampdist}. 

% In other words, the $\mathbf{w}$ term in the numerator of Equation \ref{eq:compute_gen_comp} is subtly distinct from while remaining identical to $\mathbf{w}$ in the denominator. The numerator $\mathbf{w}$ refers to the parameters involved in the conditional population model: stellar mass, sSFR, inclination, metallicity, and redshift. On the other hand, $\mathbf{w}$ in the denominator refers to the free parameters of \prosp (see \citeauthor{Leja2019} \citeyear{Leja2019} for details on these). Even the dependent variables in our model, $n$ and $\tau_2$, are not taken into account in the numerator but are free parameters in \prosp. 

% Yet at the same time, we are considering the exact same posterior samples in numerator and denominator. The caveat is that some parameters that matter in the numerator do not matter in the denominator and vice versa. However, we still need to account for all \prosp parameters when calculating the likelihood with respect to \prosp priors. After doing so, we can calculate the overall likelihood for galaxy $k$ and posterior sample $i$.

\begin{equation}
    \ln \mathcal{L}_{k,m} = \ln \mathcal{L}_{{\rm hyper},k,m} - \ln \pi_{k,m},
\end{equation}
where $\pi_{k,m}$ is the \prosp prior on the subset of variables used in our population model evaluated at the given \prosp sample, indexed by $k,m$, as described in Appendix \ref{sec:prior}.
% \textbf{***Note***: This equation doesn't account for intermediate (\prosp) priors as of yet. Can we simply use marginalized histograms of samples from \prosp priors and consider them to be independent of one another? This way, we could just subtract the log of each prior involved in the model. Another question is whether the independent variables and the dependent variable would be treated the same.}

% Then, to marginalize over samples, for each galaxy, we use the so-called ``logsumexp'' operation. In order to prevent overflows in the exponentiation operator, we temporarily change variables to make the argument of the exponential just slightly below zero (by subtracting the max log likelihood for any of the samples).

We have one more consideration for the overall likelihood calculation that is particular to our model formulation. While interpolations provide a flexible and essentially non-parametric form for the models, they often struggle with smoothness, especially when the data has strong variability or noise. This last situation applies to the \prosp data, so we add a constraint to help remove unnaturally large variations from the model. 

For the constraint, we stipulate that the differences between the dependent variables at adjacent grid points for the interpolation follow Gaussian distributions with mean $\mu=0$ and a fixed standard deviation $\sigma$. While the value of $\sigma$ could in principle be inferred as an additional model parameter, in practice this presents a challenge to the NUTS sampler. For simplicity we have instead tried several fixed values, and we find that for our 5-D models, $\sigma=1.0$ allows for a good balance between smoothness and detail. Note that the dimensions of $\sigma$ are $n$ or $\tau$ divided by the (typically logarithmic and adaptive) spacing between adjacent grid points. We sum over the log probability densities of every difference to get the total ``pair potential,'' which we denote $\gamma_\mathrm{pair}$. This quantity is \textbf{not} galaxy-dependent and is instead a constant for a particular choice of $\boldsymbol{\theta}$.

The total likelihood $\ln \mathcal{L}_{\rm tot}$ can now be calculated with the following equations. Here, $M$ is once again the total number of posterior samples per galaxy, $N$ is the total number of galaxies, and $\ln \mathcal{L}_{k,{\rm max}}$ is the maximum likelihood posterior sample for a given galaxy. This last quantity is used to prevent extreme numbers in the exponentiation process.

\begin{align}
    \ln\mathcal{L}_{k,{\rm marg}} &= \ln \mathcal{L}_{k,{\rm max}} + \ln \left[ \sum_{i=1}^M \exp{ \left( \ln{\mathcal{L}_{k,i}}-\ln \mathcal{L}_{k,{\rm max}} \right)} \right] \\
    \ln \mathcal{L}_{\rm tot} &= \gamma_\mathrm{pair} + \sum_{k=1}^N \mathcal{L}_{k,{\rm marg}}
\end{align}

\subsection{Summary of Models} \label{subsec:modsumm}

Using the posterior samples from \prosp of nearly $\sim 30,000$ 3D-HST galaxies at $0.5<z<3.0$ \citep{Leja2019,Leja2020} and the hierarchical Bayesian framework described throughout this section, we have modeled dust attenuation curves as a function of stellar mass, SFR, metallicity, redshift and axis ratio. In Appendix \ref{sec:simres}, we demonstrate how successfully our framework works by applying it to a case in which the true relation is known.

Given the connection between dust attenuation slope and normalization \citep[][and references therein]{SalimNarayanan2020}, we model the slope ($n$) and optical depth ($\tau$) of the attenuation curves jointly, with a bivariate Gaussian scatter. Going forward, we refer to such models as ``bivariate.''

We also calculate models with only $n$ as the dependent variable and $\tau$ as an independent variable. Such cases employ a univariate Gaussian error distribution rather than the more complex bivariate distribution described in \S \ref{subsec:config}. We refer to such models as ``univariate.'' We find that in general the best model for $n$ includes $\tau$ as an independent variable, as expected from \cite{SalimNarayanan2020}. 

Our models come in two flavors: single- and two-component dust models. The latter models follow directly from the dust treatment of \prosp. Diffuse dust, parameterized by Equation \ref{eq:diffdust}, affects all stars in the galaxy, whereas birth cloud dust, parameterized by Equation \ref{eq:tau1}, affects only stars under 10 Myr old. When the star formation history of a galaxy (whether observed or modeled) is not well defined or known, it is useful to have a one-component effective dust model. Once again, we use the flexible parameterization of Equation \ref{eq:diffdust}.

Therefore, we have a total of four 5-D models:
\begin{enumerate}
    \item Univariate (dependent variable $n$), two-component dust model
    \item Univariate (dependent variable $n_{\rm eff}$), one-component dust model
    \item Bivariate (dependent variables $n$ and $\tau$), two-component dust model
    \item Bivariate (dependent variables $n_{\rm eff}$ and $\tau_{\rm eff}$), one-component dust model
\end{enumerate}

All of these models are functions of log stellar mass, log SFR, log stellar metallicity, redshift, and a fifth parameter. For the univariate cases, the fifth parameter is the dust optical depth $\tau$, whereas for the bivariate cases the parameter is the axis ratio $b/a$, which is a proxy for inclination.

In the case of the two-component dust models, we model only the diffuse dust with this 5D configuration. Given the close relation between diffuse and birth cloud dust \citep[e.g.,][]{Calzetti1994,Price2014}, which is enforced in \prosp priors, as well as the difficulty to distinguish its relatively small effects on unresolved galaxy spectra, we model birth cloud dust (optical depth) as a function solely of diffuse dust (optical depth). Though this is a simpler model with only one independent variable, it is still a proper population model with the setup described in this section.

\section{Results} \label{sec:results}

In this section, we explore the relationships among dust optical depth, attenuation slope, stellar mass, SFR, metallicity, redshift, and axis ratio based on population Bayesian models. As detailed in \S \ref{subsec:modsumm}, we have four 5-D linear interpolation models for diffuse and effective dust as well as one 1-D model for birth cloud dust. 

In \S \ref{subsec:bcd}, we highlight the quasi-linear relation between diffuse dust and birth cloud dust optical depth. In \S \ref{subsec:comp}, we show that our population model produces different but more accurate trends than a simple model that ignores uncertainties in individual galaxy parameter values. Then, in \S \ref{subsec:AvS}, we compare our derived attenuation slope vs $A_V$ relation to the literature. Next, we present the relation among $\tau_{\rm eff}$, stellar mass, and SFR in \S \ref{subsec:msde}. Subsequently, in \S \ref{subsec:inclination}, we study the effects of inclination on dust attenuation. We then exhibit the cosmic evolution of attenuation in \S \ref{subsec:redshift}. Finally, we examine the intricacies of dust attenuation slope in \S \ref{subsec:univariate}.

\subsection{Birth Cloud Dust} \label{subsec:bcd}

\begin{figure}
    \centering
    \resizebox{\hsize}{!}{
    \includegraphics{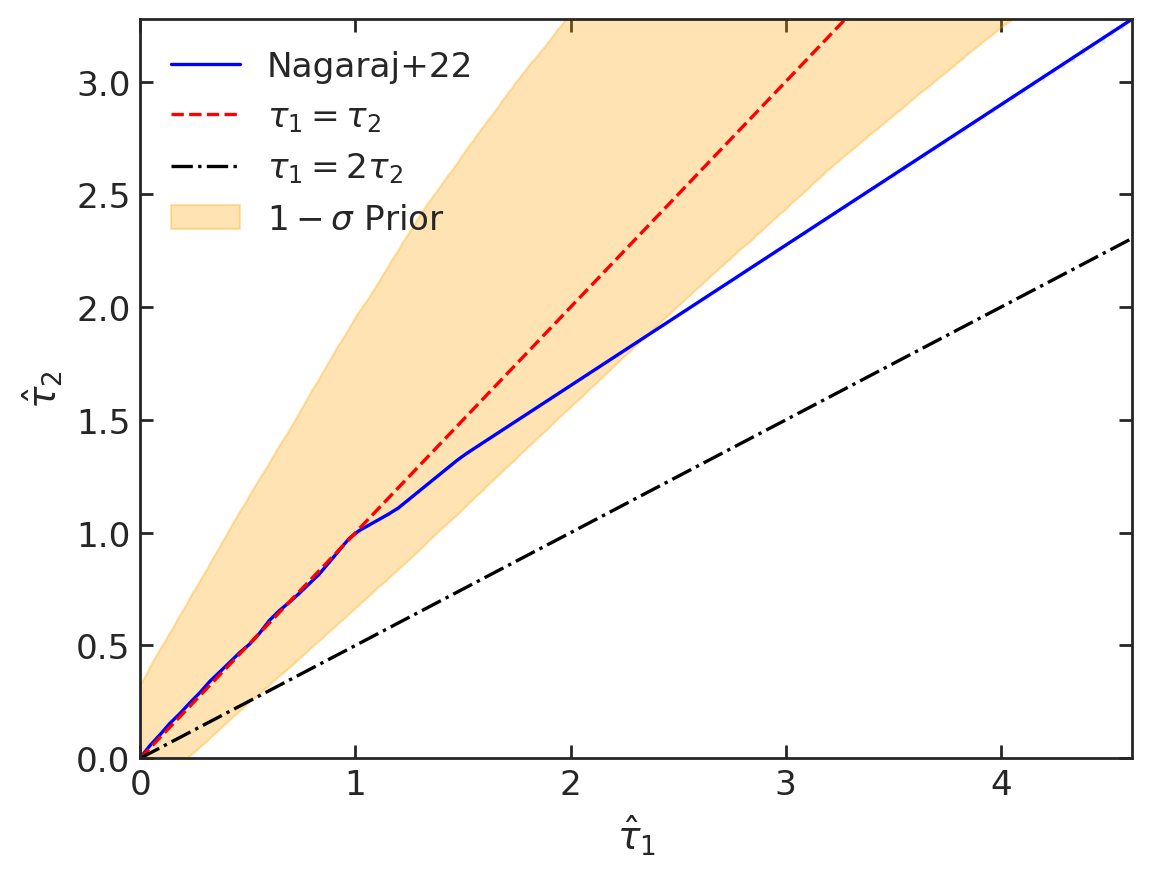}}
    \caption{Result of modeling diffuse dust optical depth $\tau_2$ solely as a function of birth cloud optical depth $\tau_1$. The model is nearly linear but does not have a slope of 1 (except when $\tau_2 \lesssim 1$). Note that uncertainties on the mean relation are shown but too small to be visible. We also show the \prosp $1-\sigma$ prior region in $\tau_2$ vs $\tau_1$ space in orange. As $\tau_1$ increases, the model moves farther from the prior, highlighting the influence of the photometry on the result. }
    \label{fig:dust1}
\end{figure}

In Figure \ref{fig:dust1}, we show our model for diffuse dust optical depth $\tau_2$ vs birth cloud dust optical depth $\tau_1$. Note that although uncertainties on the mean relation are shown on the plot, they are too small to be visible. We also show the $1-\sigma$ \prosp prior region in $\tau_2$ vs $\tau_1$ space in orange.

We can see that the relationship is close to linear, but the values are not consistent with a slope of 1 (except when $\tau_2 \lesssim 1$). Birth cloud dust tends to have a higher normalization (optical depth) than diffuse dust. Moreover, as $\tau_1$ increases, the model curve moves farther from the \prosp prior, showing that the result in Figure \ref{fig:dust1} is not merely a restatement of the \prosp prior but rather one that is informed by photometric data, and indeed qualitatively similar to results using the same data \citep{Price2014}. We discuss the physical context and implications of the birth cloud -- diffuse dust connection in \S \ref{subsec:diffbirth}.

\subsection{Benefits of the Population Approach} \label{subsec:comp}

We find that the more sophisticated population modeling approach with a proper error treatment gives a qualitatively and quantitatively different answer than a simpler one in which parameters of individual galaxies are treated as singular points. This corroborates the conclusion from \cite{Qin2022} that degeneracies between $n$, $A_V$, and the intrinsic UV slope $\beta_0$ from SED fitting strongly influence the resulting slope vs $A_V$ relation. 

\begin{figure*}
    \centering
    \resizebox{\hsize}{!}{
    \includegraphics{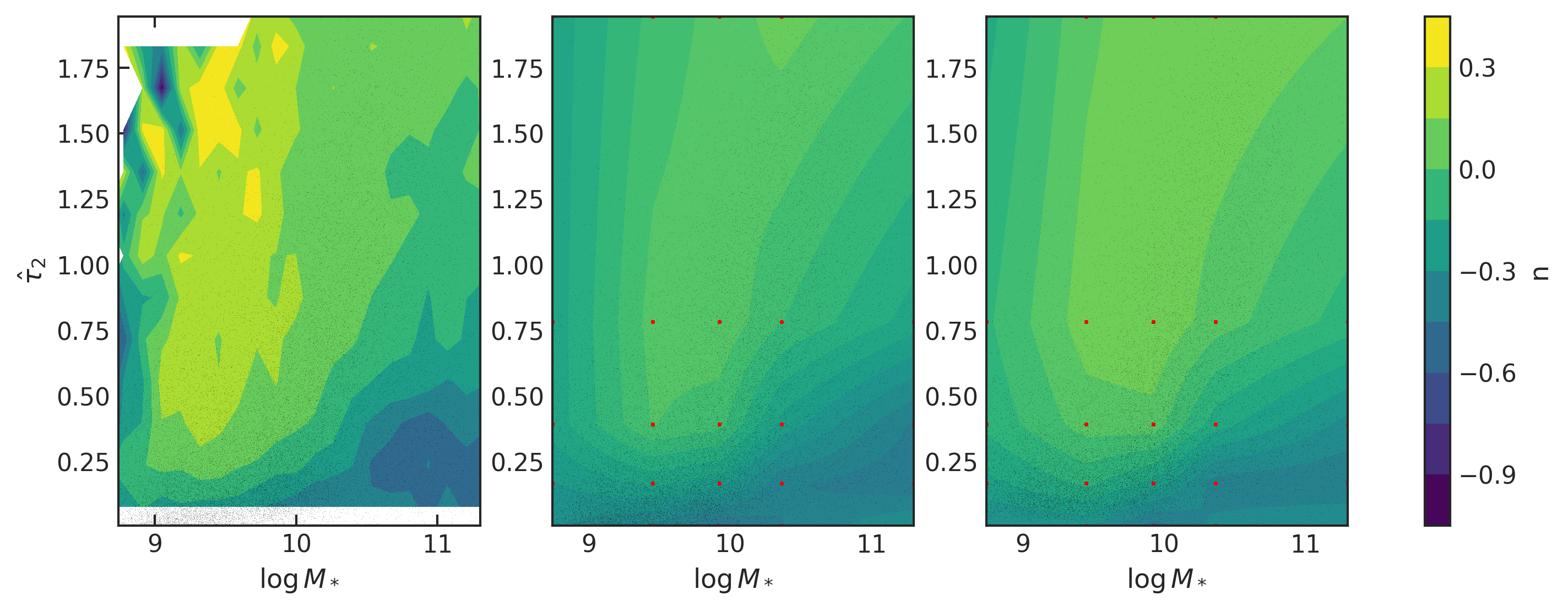}}
    \caption{Prediction of diffuse dust attenuation slope parameter $n$ vs stellar mass and $\tau_2$ using three methods. Left panel: we bin the $\tau_2$ vs $\log~M_*$ parameter space into $25 \times 25$ and record the median $n$ value from \prosp within the bin as long as at least 10 galaxies are in the bin. Maximum likelihood values for each parameter are assumed. Middle panel: we show the result of a simple Bayesian model in which we take the maximum likelihood solutions for each parameter and ignore measurement uncertainty. Right panel: result of our univariate diffuse dust model. The red points in the middle and right panels show the interpolation grid in the $\tau_2$ vs $\log~M_*$ space. From all three panels, we observe that the relation among $n$, $\tau_2$, and $\log~M_*$ is strong. We find a qualitatively and quantitatively different result in the population model (right) than the simple Bayesian model (middle), highlighting that properly accounting for uncertainties in individual galaxy parameters substantially affects the results. While it is not shown here, the population model is able to explain more of the scatter in the data than the simpler model. }
    \label{fig:comp}
\end{figure*}

In Figure \ref{fig:comp}, We show the diffuse dust slope parameter $n$ over 2-D cross-section of $\tau_2$ vs log stellar mass in three different situations. In the left panel, we take the maximum likelihood \prosp fit for each galaxy, bin the space into a $25 \times 25$ grid, and record the median $n$ value in each bin as long as at least 10 galaxies are within the bin. This is by far the simplest and most direct way of exploring the relationships between galaxies, but if the posterior volumes of individual galaxies are large (i.e., significant measurement errors), which is the case here, galaxies are likely to end up in the wrong bins. Furthermore, if posteriors are correlated, the population trends may reflect those degeneracies rather than any true trends. In addition, the maximum likelihood solution is simply a particularly interesting moment of the full parameter posterior; using the posterior directly produces more reliable results.

% Also, the binning method gets increasingly difficult as the dimensionality desired increases since either more bins (fewer galaxies per bin) or larger bins (coarseness issues) would be required for the same number of galaxies. 

In terms of identifiable features, we see that the relationship between $n$, stellar mass, and $\tau_2$ is very pronounced, with the shallowest attenuation curves at high $\tau_2$ and $\log \left(M_*/M_\odot \right) \sim 9.5$. 

In the middle panel, we show the results of a simple Bayesian model in which we continue to ignore uncertainties in the galaxy properties, just limiting ourselves to the maximum likelihood \prosp fit, which simplifies the likelihood calculations. We marginalize (see Appendix \ref{sec:marg}) over the other three parameters (log SFR, log metallicity, and redshift). 

% \footnote{We marginalize by sampling from the \prosp likelihood distribution for each quantity. More specifically, the samples are posterior samples that have been re-weighted to remove the effects of the \prosp prior distribution. We average $n$ and/or $\tau$ over various configurations of the marginal parameters}

We can see the strongest trend from the left panel reproduced in our model: the shallowest slopes are found at higher $\tau_2$ values and stellar masses around $\log \left(M_*/M_\odot \right) \sim 9.8$ whereas the attenuation curves at $\log \left(M_*/M_\odot \right) \sim 11$ and low $\tau_2$ are quite steep.

In the right panel, we show the mean results (not including the intrinsic scatter) of our univariate model for diffuse dust. The marginalization is performed in the same way as for the simple model. We find largely the same results as in the simple model, but with a slightly more pronounced shallow-slope peak for $\log~M_*\sim 10$ and $\tau_2 \gtrsim 0.5$. 

Before we analyze Figure \ref{fig:comp}, we must mention a couple of caveats. First, the left panel is not an indication of the truth as it is subject to the errors in measurements of individual galaxy parameters, though the median binning method is robust to some non-systematic sources of error. Second, the degree of similarity between the binned parameters and the models depends on the particular cross-section. For the $\tau$ vs stellar mass space (Figure \ref{fig:comp}), the population model shows a stronger trend, slightly more in accordance with the binned parameters, but for a few other cross-sections, the simple model shows stronger trends. 

Nevertheless, what we observe from Figure \ref{fig:comp} is that the simple model and population model yield qualitatively and quantitatively different results. This shows that properly accounting for covariances in the galaxy parameters affects the result. In other words, employing a population or hierarchical model will produce different and more accurate results.

The fact that the smaller-scale and more minor trends in the left panel are not reproduced by the models is not surprising. Our smoothing operation (see \S \ref{subsec:moddetails}) allows our interpolation models to converge and avoid extreme variations on small scales due to noisy data. However, it does also prevent the recovery of small scale or extremely minor trends. Of course, such trends as found directly from data (i.e., binning) are quite suspect and are often not statistically significant.

One thing not conveyed through Figure \ref{fig:comp} is the relative amounts of uncertainty in the simple vs population models. We find that the simple Bayesian model is unable to recover as much of the true trends and indeed predicts a larger intrinsic scatter ($\log \sigma=-1.21$ compared to $\log \sigma=-1.72$ in the population model). In other words, our population model is more successful at attributing causation to the input parameters than the simpler model.

% We are better able to replicate the major trends in our population model (right) than the simple Bayesian model (middle).

\subsection{Relationship between Slope and $A_V$} \label{subsec:AvS}

\begin{figure*}
    \centering
    \resizebox{\hsize}{!}{
    \includegraphics{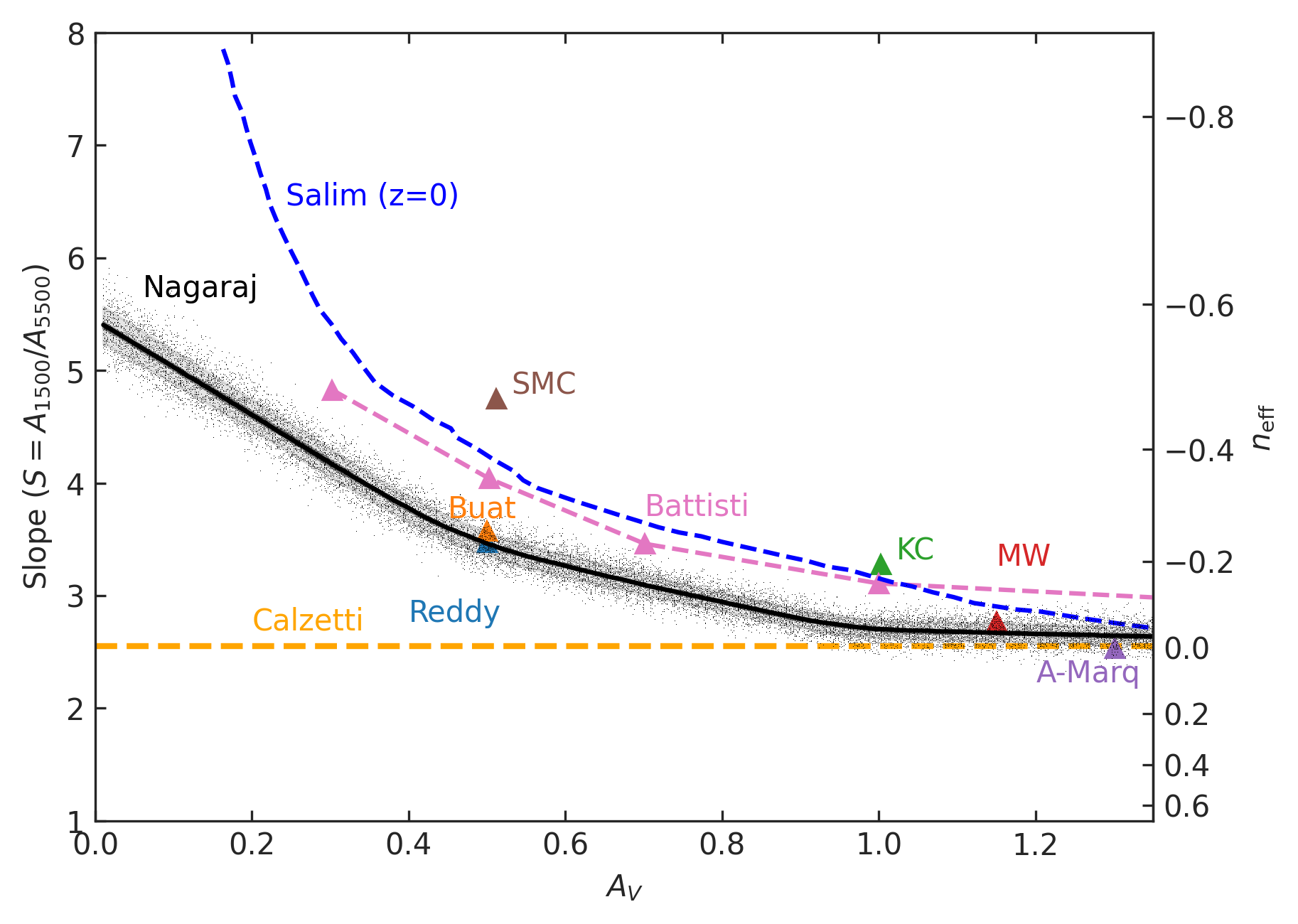}}
    \caption{Dust attenuation slope vs $A_V$ as predicted by our one-component univariate population model in comparison to previous studies: \cite{Calzetti2000}, \cite{Buat2012}, \cite{Kriek2013}, \cite{Reddy2015}, \cite{Salim2018}, \cite{Battisti2019}, \cite{AlvarezMarquez2019}. The literature values are taken from Figure 10b in \cite{SalimNarayanan2020}. Average extinction-law values for the Milky Way and Small Magellanic Cloud are included. We show the best-fit curve as a solid black line with the mean standard deviation shaded darkly and the sample standard deviation shaded lightly. We see that our model predicts a flatter relation between slope and $A_V$ compared to the literature, with the exceptions of \cite{Buat2012} and \cite{Reddy2015}. We provide a possible explanation in \S \ref{subsec:AvSDisc}. }
    \label{fig:AvSModel}
\end{figure*}

In Figure \ref{fig:AvSModel}, we show the relation between effective slope and $A_V$ for our one-component univariate dust model. The quantity $S$ is defined as $S=A_{1500}/A_V$. For convenience, we show the parameter $n$ used in our models on the right y-axis.

We compare our results to \cite{Buat2012}, \cite{Kriek2013}, \cite{Reddy2015}, \cite{Battisti2019}, \cite{AlvarezMarquez2019}, as calculated and plotted by \cite{SalimNarayanan2020}, in Figure \ref{fig:AvSModel}. We also include the results from \cite{Salim2018}, as plotted by \cite{SalimNarayanan2020}, for the local universe, and plot the \cite{Calzetti2000} relation as a baseline. For reference, the average extinction curves for the Milky Way and Small Magellanic Cloud are included. 

In the univariate model, $n_{\rm eff}$ is a direct function of $\tau_{\rm eff}$. We vary $\tau_{\rm eff}$, or more accurately $A_V \approx 1.086\tau_{\rm eff}$, on a grid from $0$ to roughly $1.4$, and marginalize over the other four independent variables.

We randomly sample from the model posterior and apply each instance (a single posterior sample per galaxy) to all ``galaxies'' to create a curve per sample set. We take the average curve (black line) and calculate the standard deviation of the average curve (dark shaded region) and of all realizations of the model posterior in general (light shaded region). 

Our average curve tends to have a less pronounced (shallower) relation between slope and optical depth than most previous results, with the notable exceptions of \cite{Buat2012} and \cite{Reddy2015}. We discuss the implications of our results in \S \ref{subsec:AvSDisc}.

\subsection{Dust Optical Depth vs Stellar Mass and SFR} \label{subsec:msde}

\begin{figure*}
    \centering
    \resizebox{\hsize}{!}{
    \includegraphics{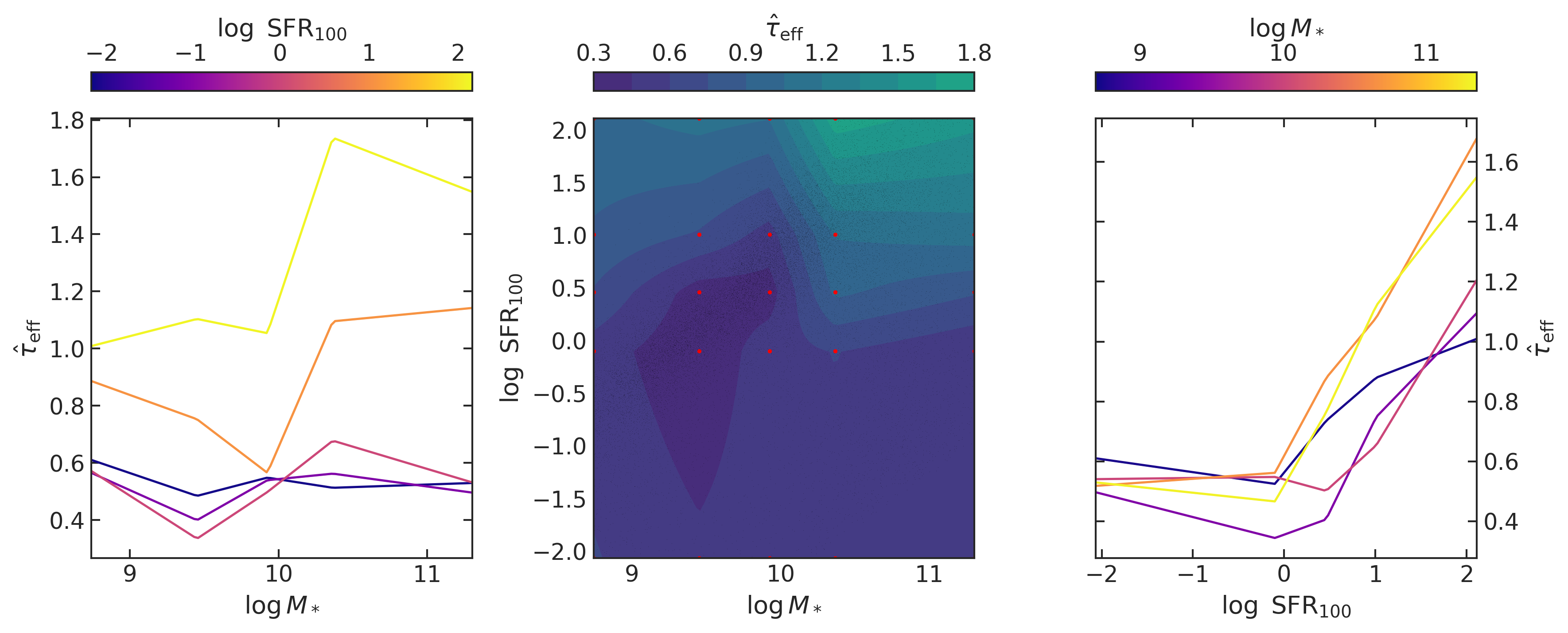}}
    \caption{Effective dust optical depth $\tau_{\rm eff}$ as a function of stellar mass and SFR, marginalized over $\log~(Z/Z_\odot)$, $z$, and $b/a$. The center panel shows the full 2-D cross-section while the left and right panels show the 1-D cross sections at five values of the other variable. The red points represent the interpolation grid while the faint black pixels show the \prosp mass-complete sample. We observe that $\tau_{\rm eff}$ is nearly constant at low SFRs, while it increases with stellar mass in vigorously star forming galaxies until $\log~M_*\sim 10.5$, above which it decreases slightly. $\tau_{\rm eff}$ increases with SFR at all stellar masses when $\log~{\rm SFR}\gtrsim 0$.}
    \label{fig:cmp_msde}
\end{figure*}

We find that the dust effective (and diffuse) optical depth $\tau$ depends most strongly on SFR. The dependence on the other four parameters is not as significant but still substantial. In Figure \ref{fig:cmp_msde}, we show the relation between $\tau_{\rm eff}$, SFR, and stellar mass. 

The center panel shows the 2-D cross-section of the model marginalized over the other three parameters. The behavior of the model is quite complex, making it quite clear that the dependence of $\tau$ on stellar mass and SFR cannot be separated. In other words, dust optical depths in massive galaxies are generally higher at a fixed SFR than in low-mass galaxies.

% Once again, the graininess comes from a combination of the marginalization and the intrinsic scatter in the relation.

The left and right panels show the 1-D cross-sections of the model at five fixed values of the other variable. The left panel shows $\tau_{\rm eff}$ vs log stellar mass. We observe that at the three lowest values of SFR, the relation in stellar mass is nearly flat, suggesting little to no correlation between $\tau_{\rm eff}$ and stellar mass in galaxies with low star formation rates. On the other hand, $\tau_{\rm eff}$ clearly increases with stellar mass for star-forming galaxies for $\log~M_*\lesssim 10.5$. For $\log~M_*\gtrsim 10.5$, $\tau_{\rm eff}$ seems to slowly decrease with stellar mass for most values of SFR.

Meanwhile, the right panel shows $\tau_{\rm eff}$ vs log SFR. Interestingly, for all five stellar masses, $\tau_{\rm eff}$ stays nearly constant until $\log~{\rm SFR}\sim 0$, above which it quickly increases.

\subsection{The Effects of Inclination on Dust Attenuation} \label{subsec:inclination}

In our bivariate model, we include the axis ratio $b/a$ as a proxy for inclination. As mentioned in \S \ref{sec:data}, the axis ratio data comes from \cite{vanderWel2014} through \texttt{GALFIT} \citep{Peng2002,Peng2010} fits. 

\begin{figure*}
    \centering
    \resizebox{\hsize}{!}{
    \includegraphics{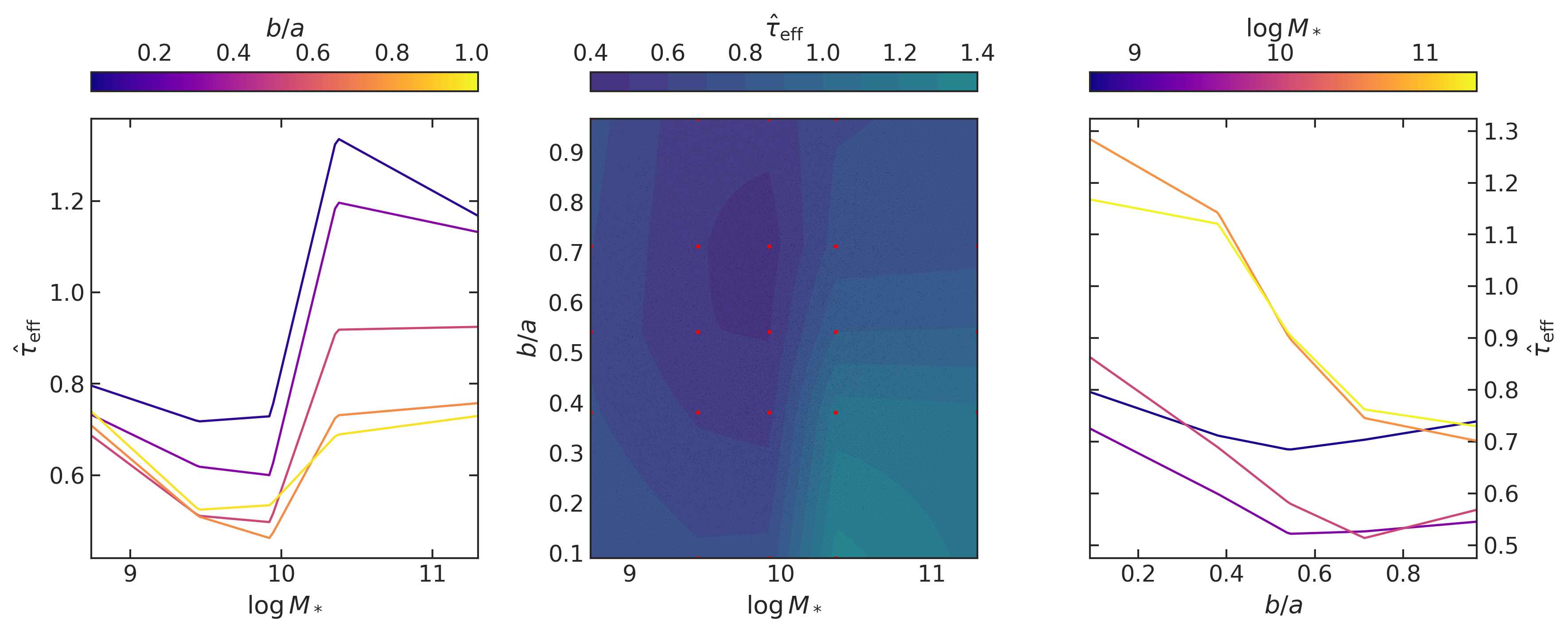}}
    \caption{Effective dust optical depth $\tau_{\rm eff}$ vs $\log~M_*$ (left) and axis ratio $b/a$ (right), marginalized over $\log$ SFR, $\log~(Z/Z_\odot)$, and $z$. The center panel shows the full relationship while the two side plots show the 1-D cross-sections for five fixed values of the other parameter. We find that edge-on galaxies tend to have larger optical depths than face-on galaxies, but only for galaxies with $\log~M_* \gtrsim 10$. Low mass galaxies in the distant universe have still not developed a distinct disk-like morphology, resulting in no relationship between $b/a$ and viewing $\tau_{\rm eff}$.}
    \label{fig:cmp_mide}
\end{figure*}

In our model, $b/a$ is roughly as important as stellar mass in determining the dust optical depth. Figure \ref{fig:cmp_mide} shows $\tau_{\rm eff}$ over the 2-D cross-section of $b/a$ vs $\log~M_*$ (center) in addition to the 1-D cross-sections of $b/a$ (right) and $\log~M_*$ (left). 

% From the 1-D cross-sections, we observe that $\tau_{\rm eff}$ is nearly constant with $b/a$ for masses $\log~M_* \lesssim 10$. 

In the $\tau_{\rm eff}$ vs $\log~M_*$ plot, we can see that for $\log~M_* \lesssim 10$, the curves at all values of $b/a$ are overlapping. As mass increases, the curves spread out, and we can see that edge-on galaxies (represented by the $b/a=0.09$ curve) have higher $\tau_{\rm eff}$ values than the face-on galaxies (represented by the $b/a=0.97$ curve). 

Similarly, in the $\tau_{\rm eff}$ vs $b/a$ plot, we see that for the highest two $\log~M_*$ values, $\tau_{\rm eff}$ decreases as we move toward face-on galaxies. On the other hand, the curves for the lowest three $\log~M_*$ values are closer to constant.

\begin{figure*}
    \centering
    \includegraphics[scale=0.75]{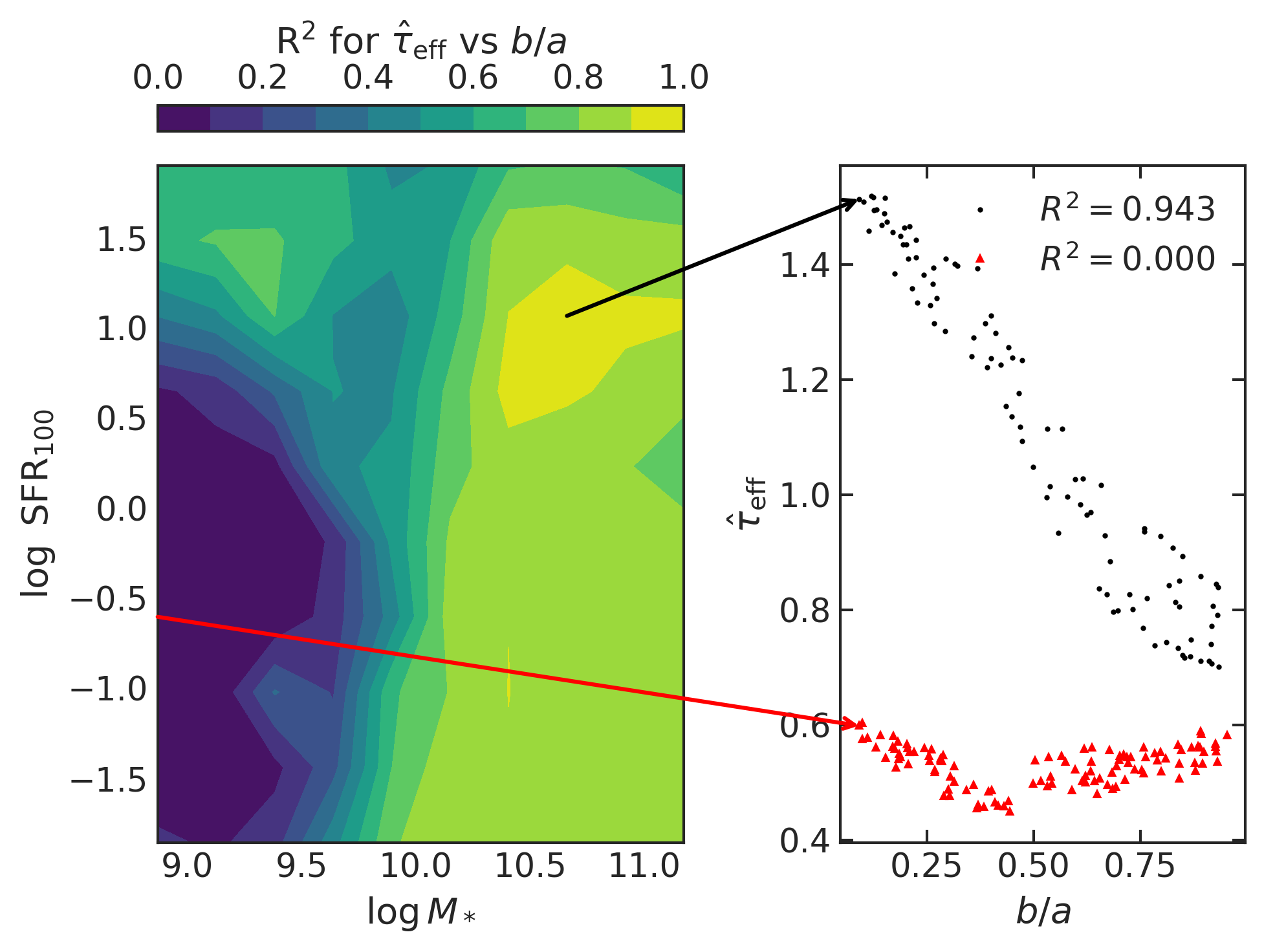}
    \includegraphics[scale=0.75]{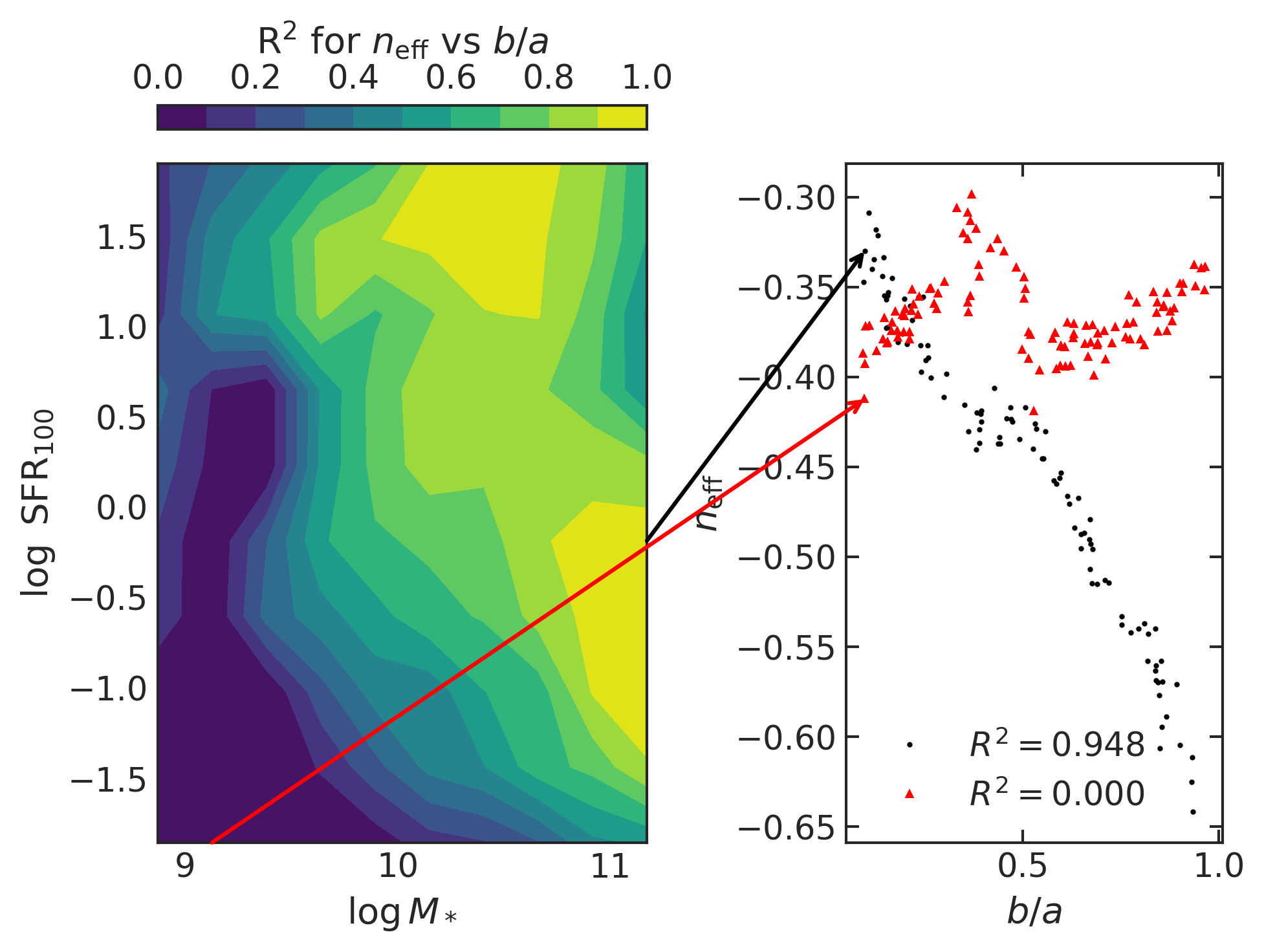}
    \caption{The left panels show the correlation ($R^2$) between model $\tau_{\rm eff}$ (top) and $n_{\rm eff}$ (bottom) vs axis ratio $b/a$ as a function of stellar mass and SFR. The right panels show the modeled relation between $\tau_{\rm eff}$ (top) or $n_{\rm eff}$ (bottom) and $b/a$ at the locations in parameter space with the highest and lowest $R^2$ values, as indicated by the black and red arrows, respectively. We find that inclination is most important for dust attenuation (at least the optical depth parameter) for massive galaxies, especially those with a high SFR. As expected from Figure \ref{fig:cmp_mide} and our physical understanding (\S \ref{subsec:incdisc}), for such galaxies $\tau_{\rm eff}$ decreases as we move toward face-on orientations. We also indirectly observe the strong connection between $\tau_{\rm eff}$ and $n$ (Figure \ref{fig:AvSModel}) in the bottom right panel with the steepening of the attenuation curve as we move toward face-on orientations. In parallel, we see that inclination has little impact on dust attenuation curves in low-mass, low-SFR galaxies, further supporting the picture that high-redshift, low-mass galaxies lack disk-like morphologies.}
    \label{fig:rsq_i}
\end{figure*}

To probe the question of inclination farther, we show how our modeled $\tau_{\rm eff}$ and $n_{\rm eff}$ correlate with $b/a$ as a function of $\log~M_*$ and $\log$ SFR in Figure \ref{fig:rsq_i}. The left panels of the plots show the $R^2$ value of the model $\tau_{\rm eff}$ (top) and $n_{\rm eff}$ (bottom) vs $b/a$ in 225 bins in the $\log$ SFR vs $\log~M_*$ parameter space. In other words, we are determining how important inclination is in the dust attenuation model for different values of stellar mass and SFR. In the right panels, we show the predicted model for $\tau_{\rm eff}$ (top) and $n_{\rm eff}$ (bottom) vs $b/a$ at the highest and lowest $R^2$ values in the parameter space. 

We see that inclination plays the biggest role in determining dust attenuation curves for massive, vigorously-star-forming galaxies. For $n_{\rm eff}$, the peak in $R^2$ is at slightly lower SFRs than for $\tau_{\rm eff}$, but the overall effect on the attenuation curve is as stated. As expected from Figure \ref{fig:cmp_mide}, for massive, star-forming galaxies, $\tau_{\rm eff}$ decreases as we move toward face-on galaxies. Given the strong relation between $\tau_{\rm eff}$ and $n_{\rm eff}$ (Figure \ref{fig:AvSModel}), the steepening of the attenuation curve as we move toward face-on galaxies is a natural consequence. 

The only region where inclination plays a minimal role in attenuation is at low masses and low SFRs, which is consistent with our findings from Figure \ref{fig:cmp_mide}. In this region, we observe that $\tau_{\rm eff}$ is nearly constant while $n_{\rm eff}$ has a quasi-sinusoidal relation with $b/a$. The net effect on the overall dust attenuation curve is small. We discuss the physical implications of these findings in \S \ref{subsec:incdisc}.

\subsection{Redshift Evolution of Dust Attenuation} \label{subsec:redshift}
\begin{figure*}
    \centering
    \includegraphics[scale=0.75]{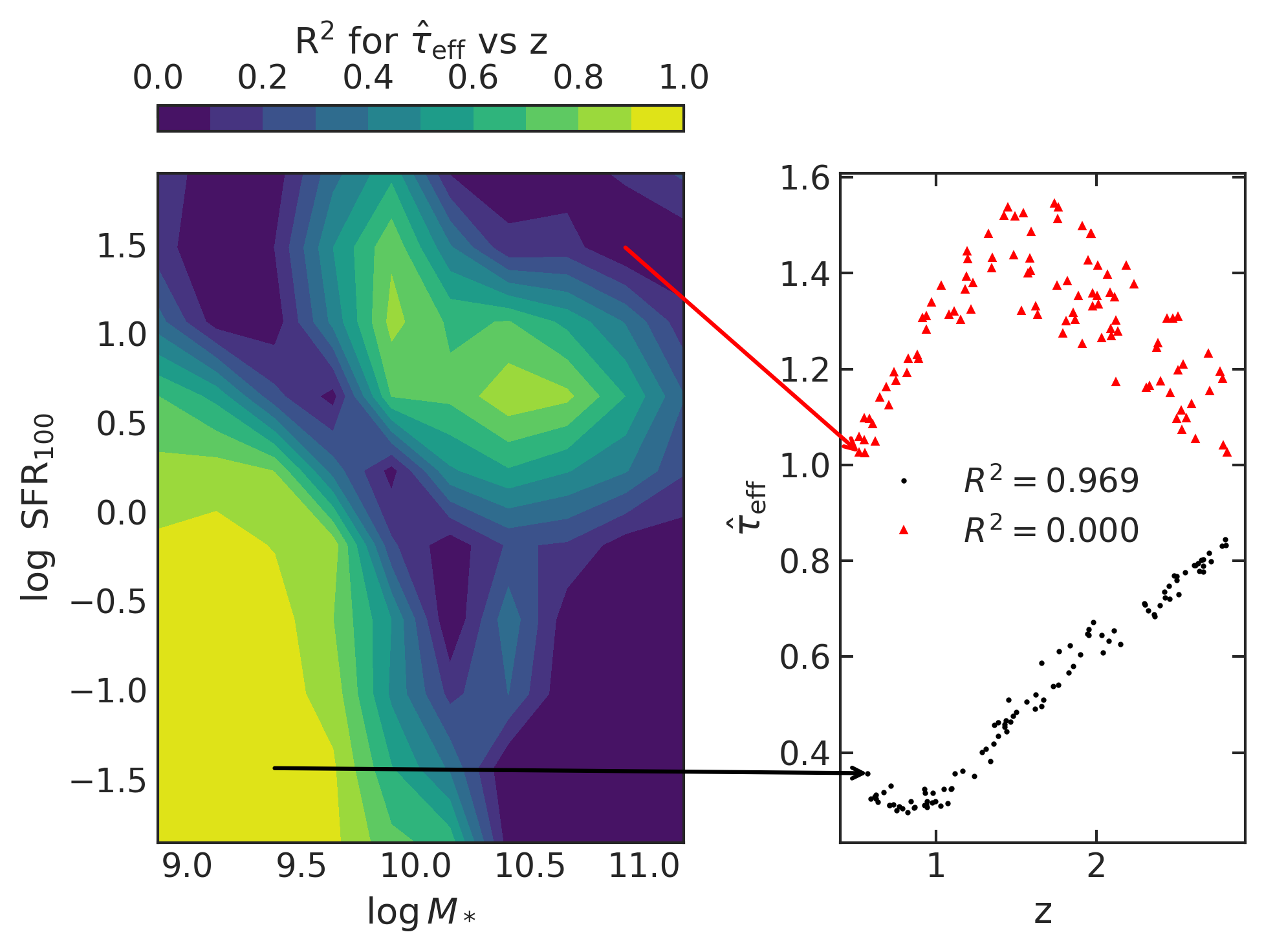}
    \includegraphics[scale=0.75]{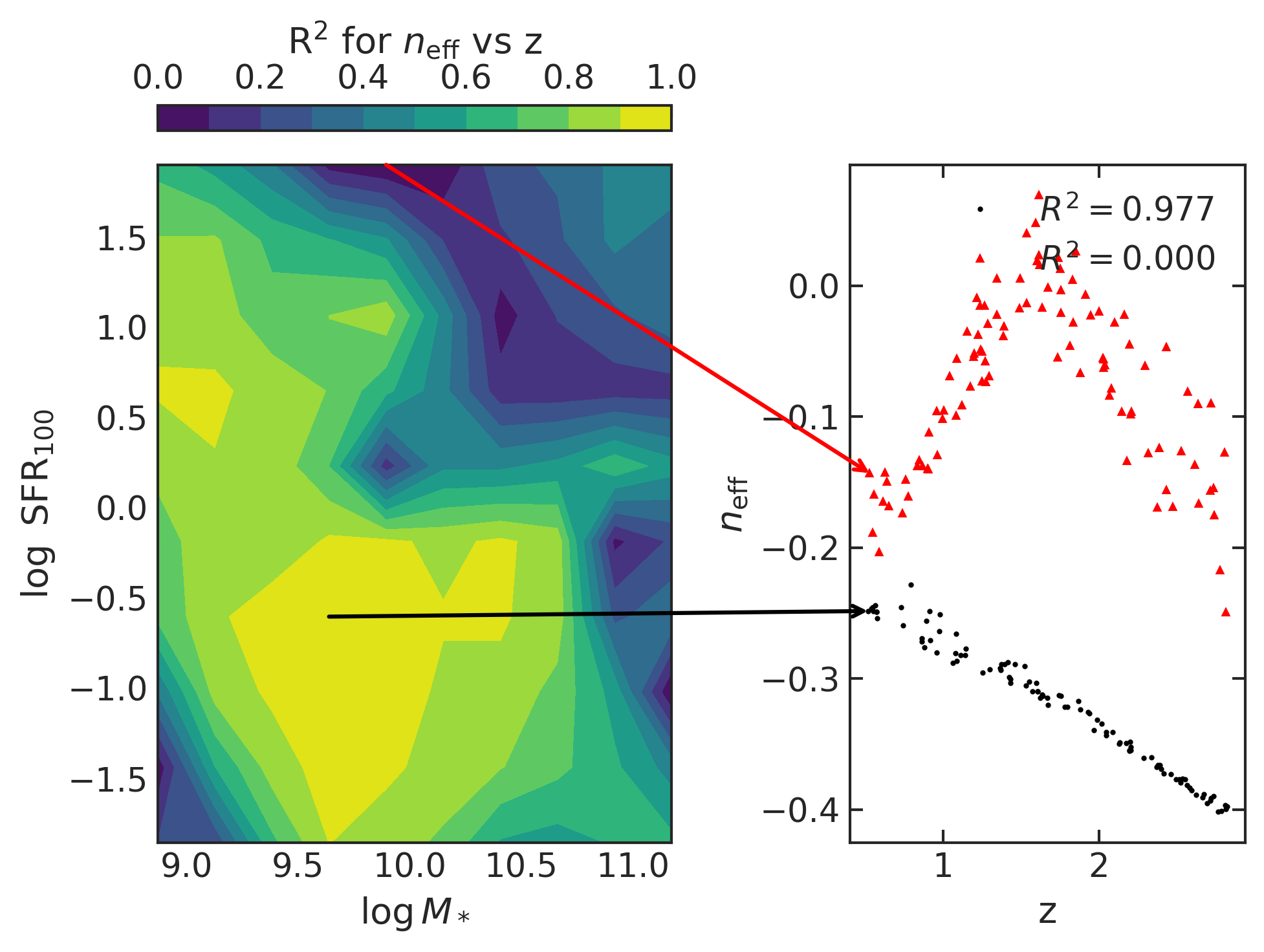}
    \caption{The left panels show the correlation ($R^2$) between model $\tau_{\rm eff}$ (top) and $n_{\rm eff}$ (bottom) vs $z$ as a function of stellar mass and SFR. The right panels show the modeled relation between $\tau_{\rm eff}$ (top) or $n_{\rm eff}$ (bottom) and redshift at the locations in parameter space with the highest and lowest $R^2$ values, as indicated by the black and red arrows. We find that the dust attenuation curve undergoes the strongest redshift evolution for low-mass, low-SFR galaxies. For such galaxies, dust attenuation curves have higher optical depths and become steeper with increasing redshift. However, the picture may be convoluted by the differential mass-completeness limit in our sample: at $z=2.1$ and $z=3.0$, the limit is $\log~M_*=9.79$ and $\log~M_*=10.15$, respectively. Therefore, for $z\gtrsim 2$ the dust evolution must be confirmed by studies that push down to lower masses.}
    \label{fig:rsq_biv_z}
\end{figure*}

In all 1-D and 2-D cross-sections of parameter space, there are no strong relations between dust attenuation curves and redshift. However, the situation changes in 3-D, as shown in Figure \ref{fig:rsq_biv_z}. On the left we plot the correlation ($R^2$) between the model parameters $\tau_{\rm eff}$ (top) or $n_{\rm eff}$ (bottom) vs redshift as a function of stellar mass and redshift. The setup of these panels is identical to those in Figure \ref{fig:rsq_i}. 

The right panels show the model result for the dust attenuation parameters $\tau_{\rm eff}$ (top) and $n_{\rm eff}$ (bottom) as a function of redshift given the stellar mass and SFR values where the highest and lowest $R^2$ values are recorded, as indicated by the colored arrows. 

We find that redshift is highly influential for galaxies with $\log~M_* \lesssim 9.7$ and $\log~{\rm SFR}\lesssim 0.5$: low-mass, low-SFR galaxies. It suggests that the dust in such galaxies evolves the most over cosmic time. From the right panels, we observe that for the low-mass, low-SFR galaxies, dust attenuation curves tend to have higher optical depths and become steeper with increasing redshift, but only for $z\gtrsim 1$. 

However, we can be certain about these results only for redshifts $0.5<z\lesssim 2$, as our mass-complete sample does not extend to such low stellar masses at the highest redshifts. To put it in quantitative terms, at $z=2.1$ and $z=3.0$, the minimum mass in our samples is $\log~M_*=9.79$ and $\log~M_*=10.15$, respectively. The differential mass-complete range may contribute to the seeming importance of redshift to the relation or at least prevent the creation of a sensible model from lack of data. Pushing down to lower masses at higher redshift, whether through better data or a more sophisticated treatment of the selection function, may change the model predictions at low stellar mass and SFR.

% The relation is stronger for $\tau_{\rm eff}$ than it is for $n_{\rm eff}$, but the former is more important in determining the overall dust attenuation curve.

% The strong redshift evolution of such galaxies may be linked to the peculiar traits of those objects at low redshift (high supernova/stellar feedback, ram pressure stripping, etc.) compared to at high redshift.

\subsection{High-level correlations between $n_{\rm eff}$, $\tau_{\rm eff}$, Stellar Mass, SFR, and Metallicity} \label{subsec:univariate}

\begin{figure*}
    \centering
    \resizebox{\hsize}{!}{
    \includegraphics{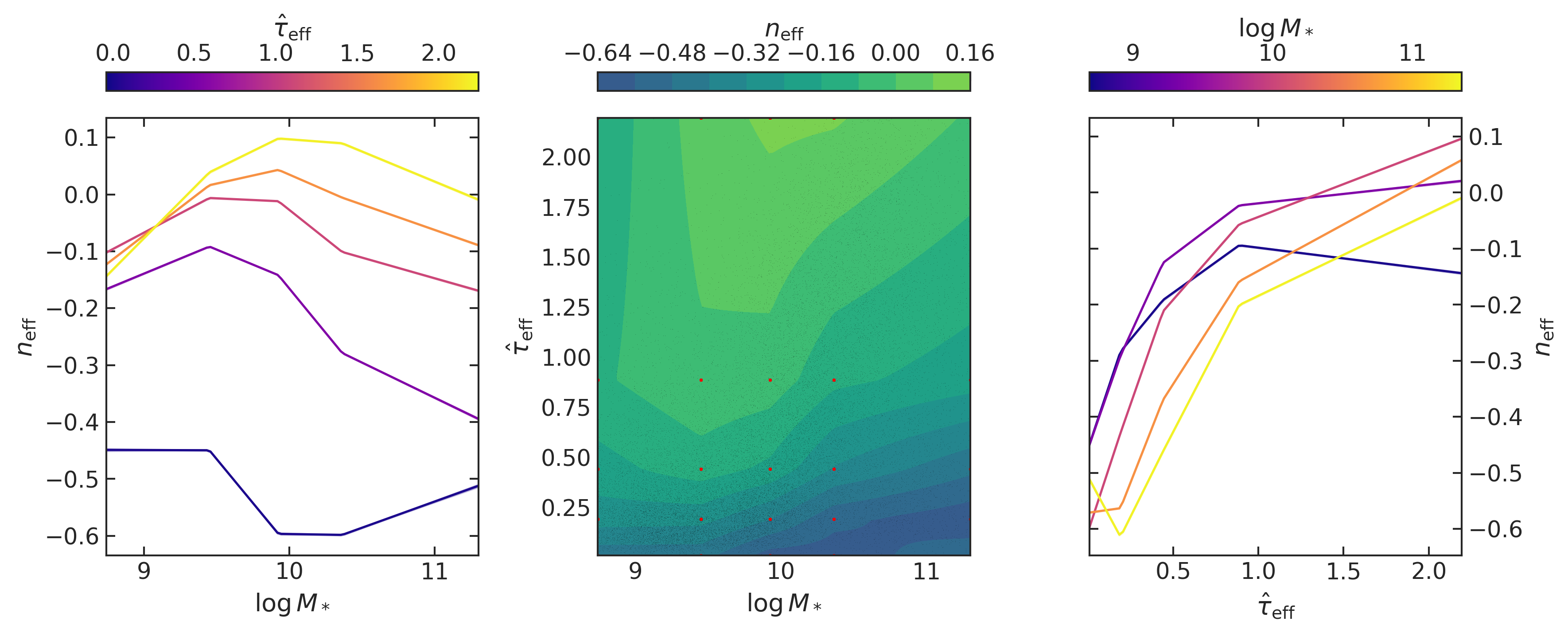}}
    \caption{Effective attenuation slope $n_{\rm eff}$ as a function of $\log~M_*$ and $\tau_{\rm eff}$, marginalized over $\log$ SFR, $\log~(Z/Z_\odot)$, and $z$. The center panel shows the full relation, while the left and right panels show the 1-D cross-sections for five values of the other parameter. From the right panel, we once again confirm the result that increasing $\tau$ leads to a shallower attenuation curve. From the left panel we see that the dust attenuation curve first gets shallower and then steeper with stellar mass. This can be explained by the relative amounts of obscured and unobscured stars of different masses and dust clumpiness. The full plausible scenario is presented in \S \ref{subsec:effslope}.}
    \label{fig:cmp_mdene}
\end{figure*}

In Figure \ref{fig:cmp_mdene}, we show $n_{\rm eff}$ as a function of $\tau_{\rm eff}$ and $\log~M_*$. Once again, the center panel shows the full function (marginalized over $\log$ SFR, $\log~(Z/Z_\odot)$, and $z$), while the left and right panels show 1-D cross-sections with $\log~M_*$ and $\tau_{\rm eff}$, respectively. From the $n_{\rm eff}$ vs $\tau_{\rm eff}$ plot, we can very clearly see that as $\tau_{\rm eff}$ increases, $n_{\rm eff}$ also increases for all stellar masses (with a slight decrease at high $\tau_{\rm eff}$ when $\log~M_* \sim 9$), which is consistent with the relationship between $A_V$ and $S$ in Figure \ref{fig:AvSModel}.

The relationship between $n_{\rm eff}$ and $\log~M_*$ is much more complex and is highly dependent on $\tau_{\rm eff}$, but in general, we see that $n$ increases (becomes shallower) and then decreases (becomes steeper) with stellar mass. 

% We discuss plausible scenarios for this finding in \S \ref{subsec:effslope}.
% We ignore the $\tau_{\rm eff}=0.01$ curve as for such a negligible optical depth the concept of slope is ill-defined. 

\begin{figure*}
    \centering
    \includegraphics[scale=0.75]{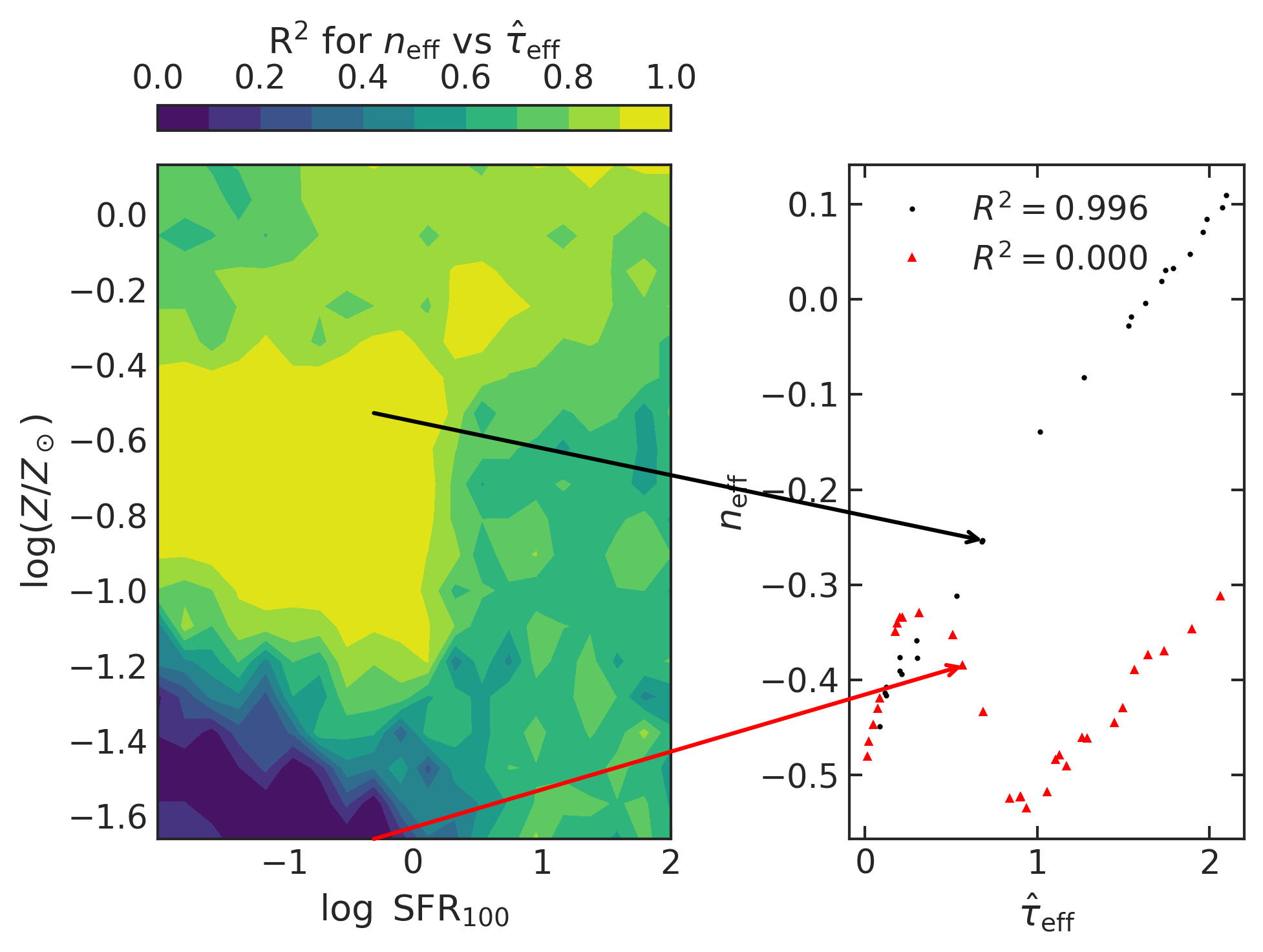}
    \caption{Left: correlation ($R^2$) between $n_{\rm eff}$ and $\tau_{\rm eff}$ as a function of stellar metallicity and SFR. Right: modeled relation between $n_{\rm eff}$ and $\tau_{\rm eff}$ at the locations in parameter space with the highest and lowest $R^2$ values, as indicated by the black and red arrows. The relationship at the highest $R^2$ location is exactly what we expect: the attenuation curve becomes flatter as we increase optical depth. On the other hand, we observe that $\tau_{\rm eff}$ is uncorrelated with $n_{\rm eff}$ when $\log~{\rm SFR}\lesssim -0.5$ and $\log~(Z/Z_\odot)\lesssim -1.2$. Specifically, the modeled relation in this regime is quasi-sinusoidal with an amplitude small compared to the range of $n$ (under $10\%$). This suggests that dust attenuation at the low-metallicity, low-SFR regime behaves differently, although a strong statement cannot be made with just the evidence we have available.}
    \label{fig:rsq_univ_de}
\end{figure*}

Next, in Figure \ref{fig:rsq_univ_de}, we show the correlation ($R^2$) of $n_{\rm eff}$ vs $\tau_{\rm eff}$ as a function of stellar metallicity and SFR (left). In the right panel we show the modeled relation between $n_{\rm eff}$ and $\tau_{\rm eff}$ at the highest and lowest $R^2$ values. Interestingly, even though we have repeatedly seen the dependence of $n$ on $\tau$ (which is still the case for the highest $R^2$ location), from this plot we observe that the dependence is not universal. Indeed, $\tau_{\rm eff}$ has a small but complex and non-linear effect on $n_{\rm eff}$ when $\log~{\rm SFR}\lesssim 0.5$ and $\log~(Z/Z_\odot)\lesssim -0.4$.

\section{Discussion} \label{sec:disc}

In this section, we discuss the scientific implications of a few key results from \S \ref{sec:results}. In \S \ref{subsec:diffbirth}, we connect the modeled relationship between diffuse and birth cloud dust optical depth to a simple physical picture depicted by \cite{Reddy2015}. Then, in \S \ref{subsec:AvSDisc}, we discuss the origins of the $S$ vs $A_V$ relation as well as why our $S$ vs $A_V$ curve is shallower than that predicted by the majority of the literature. Next, we consider the relationship between inclination and dust attenuation over a wide range of galaxies in \S \ref{subsec:incdisc}. Finally, in \S \ref{subsec:effslope}, we analyze the complex relationships between $n_{\rm eff}$ and the other galaxy properties.

\subsection{Diffuse vs Birth Cloud Dust: A Closer Look} \label{subsec:diffbirth}

In Figure \ref{fig:dust1} we show the relation between diffuse and birth cloud dust optical depth. As noted in \S \ref{sec:results}, the relation is nearly linear but inconsistent with a slope of unity: as the optical depth increases, birth cloud dust grows faster than diffuse dust. 

This can be explained by the connection between dust attenuation and SFR: the optical depth tends to increase with SFR \citep[e.g.,][]{Buat2009,Reddy2010,Reddy2015}. \cite{Reddy2015} suggest a simple model to explain the trend found in Figure \ref{fig:dust1}: as the SFR increases, dust enrichment in birth clouds is more prominent than in the diffuse ISM, meaning that on average massive stars face more dust attenuation. We refer the reader to their \S 4.3 for a more thorough discussion on this point (e.g., Figure 19).

\subsection{Relationship between $A_V$ and Slope: Origin and Literature Comparison}\label{subsec:AvSDisc}

Theory and empirical evidence suggests that the dust attenuation slope becomes shallower with increasing optical depth \citep[][and references therein]{SalimNarayanan2020}. The physical explanation, originating from \cite{Chevallard2013}, is related to the fact that redder light scatters more isotropically than bluer light.

At low optical depth, a significant fraction of the galaxy integrated light we observe arises from the equatorial plane of the galaxy, where blue light has a harder time escaping than redder light, leading to a steeper dust attenuation law. At high optical depth, we are observing the $\tau=1$ surface, which includes more light outside the equatorial plane and is therefore less dependent on the scattering differences. Also, massive stars in the $\tau=1$ vicinity may or may not be obscured by dust, leading to a shallower attenuation slope. 

% Furthermore, theory suggests that as the total dust content increases, the dust should become clumpier \citep[][and references therein]{SalimNarayanan2020}, again suggesting that some stars will be relatively unobscured while others are heavily obscured. Given that the luminosity-mass relation for stars is steeper than the initial mass function, the net effect of the clumpiness will be that UV light is dominated by unobscured massive stars, which will lead to a flattening of the attenuation curve.

In Figure \ref{fig:AvSModel}, we show effective slope vs $A_V$ for our univariate model. As mentioned in \S \ref{subsec:AvS}, our results are consistent with the expectations mentioned above that dust attenuation curves become flatter as the optical depth increases. However, our curve is shallower than the majority of literature results, with the exceptions of \cite{Buat2012} and \cite{Reddy2015}. 

\cite{Qin2022} suggest that the $S$ vs $A_V$ relation is influenced by degeneracies between parameters. As our hierarchical framework properly treats correlated parameters, one may suspect that the difference between our curve and the rest of literature originates in the statistical approach. However, we find that even using \prosp maximum likelihood parametric values directly, a very similar result (albeit with larger uncertainty) is derived. As the difference in statistical modeling is ruled out as an explanation, we move toward differences in SED modeling.

In that vein, a plausible explanation for the shallower curve is that, as mentioned in the introduction, \prosp measured higher stellar masses and lower current SFRs in comparison to previous studies \citep{Leja2019}, lessening the influence of young stars and birth cloud dust. In other words, the stars dominating the UV spectrum are slightly older and redder on average and/or are surrounded by less birth cloud dust, leading to a shallower effective dust attenuation curve.

\subsection{The Role of Inclination in Regulating Dust Attenuation} \label{subsec:incdisc}

For disk galaxies, inclination naturally plays a role in the observed dust attenuation curve: edge-on lines of sight will be subject to significantly larger dust optical depths than face-on sight lines. \cite{Chevallard2013} found that in theoretical models, dust geometry and galaxy orientation played a major role in the resulting dust attenuation curves. \cite{Doore2021} and \cite{Zuckerman2021} have found observational evidence for the importance of inclination in determining dust attenuation. Like in \cite{Zuckerman2021}, we use the more-easily-measurable axis ratio $b/a$ as a proxy for inclination.

In \S \ref{subsec:inclination}, we show that our results are in agreement with expectations and the previous results mentioned for massive galaxies. In Figures \ref{fig:cmp_mide} and \ref{fig:rsq_i}, we observe that optical depth increases significantly ($0.8$ to $1.4$) from face-on to edge-on galaxies (or more accurately, $b/a=1$ to $b/a=0$), for galaxies with $\log~M_*\gtrsim 10.5$. In this regime, the effects of inclination are roughly equal to those of stellar mass (over the whole range of masses) on the dust attenuation curve.

At the same time, the lack of relation between $\tau_{\rm eff}$ and $b/a$ for lower stellar masses is quite simple: in the high-redshift universe, low-mass galaxies have not yet formed a disk-like shape \citep{vanderWel2014}, thus obfuscating the relation between $b/a$ and viewing inclination. At higher mass, especially for star-forming galaxies, the disk structure is more pronounced, and thus the paradigm of edge-on vs face-on galaxies is established and the importance of viewing angle comes into play.

Furthermore, the lack of distinct morphological features for high-redshift, low-mass galaxies, especially with low SFRs explains the model prediction in Figure \ref{fig:rsq_i} that inclination has the least impact on dust attenuation for low-mass, low-SFR galaxies.

\subsection{Interpreting the Effective Slope} \label{subsec:effslope}

In \S \ref{subsec:AvS}, we showed that the effective attenuation slope $n_{\rm eff}$ becomes shallower as the optical depth $\tau_{\rm eff}$ increases, although the relation we find is slightly flatter than that found by previous studies. In this section, we discuss further the properties of the slope in our population model.

The slope determines the wavelength dependence of dust attenuation, thus playing an important role in the observed galaxy color. The reasons for different slopes are complex and may include star-dust geometry considerations and even chemical composition and size distribution of the dust particles themselves.

% In agreement with the literature, we find that $n$ depends most strongly on $\tau$. From a utility standpoint, this is rather unfortunate as the situations where one knows or has estimates for $\tau$ but not $n$ are limited. Nevertheless, from an astrophysical perspective, any strong relationship that can be found is of great interest.

Figure \ref{fig:cmp_mdene} shows $n_{\rm eff}$ vs $\tau_{\rm eff}$ and stellar mass. The $n_{\rm eff}$ vs $\tau_{\rm eff}$ cross-section simply reflects what we already showed in Figure \ref{fig:AvSModel}: dust attenuation curves get shallower as the optical depth increases. However, the $n_{\rm eff}$ vs stellar mass cross-section is more complex: the attenuation curve first gets shallower and then steeper with stellar mass, with the turning point becoming shallower and occurring at higher stellar mass as $\tau$ increases. Theoretical studies will need to be conducted to better understand this phenomenon.

In Figure \ref{fig:rsq_univ_de}, we show that $\tau$ has a small but complex effect on $n$ for low-metallicity, low-SFR galaxies. One implication from this result is that dust attenuation quite possibly behaves differently in the low-metallicity, low-SFR regime. Such a finding is not surprising as chemical composition of dust as well as production and destruction pathways likely depend on metallicity. In addition, most of the dust in low SFR galaxies was likely created in previous epochs, which may change the overall picture. 

However, caution must be used before making too general a statement. The low-SFR, low-metallicity regime is not well represented in the 3D-HST catalog, and metallicities are difficult to measure from SED fitting codes. Long-exposure spectroscopy focusing on stellar metallicity indicators such as the absorption line rest-frame equivalent width between $1496$ and $1506 \AA$ \citep{Sommariva2012} can help better pinpoint metallicities of galaxies in the sample and thus better inform a dust-metallicity study.

It may also be useful to study the relation between dust attenuation and gas-phase metallicity given, for example, the correlation in the local universe between polycyclic aromatic hydrocarbons (PAHs)---dust molecules containing up to 15-20\% of all carbon in dust \citep{Draine2003,Zubko2004}---and gas-phase metallicity \citep[e.g.,][]{Engelbracht2005,Madden2006}. A study using strong line indicators \citep[e.g.,][]{Kewley2008} or better yet \OIIIsimp $\lambda 4363$ (if not too faint) to measure gas-phase metallicity would help investigate the connection between dust attenuation and gas-phase metallicity. 

\section{Code for Users} \label{sec:code}

By coupling the posterior samples of our four population models (\S \ref{sec:results}) with a code that parses them and calculates dust attenuation curves, we have created the Python-based module \texttt{DustE}\footnote{https://github.com/Astropianist/DustE}. 

Users supply values for stellar mass, SFR, stellar metallicity, redshift, axis ratio, and/or dust optical depth and select which model to be used. Any variables not provided are marginalized over accordingly. Functions are provided to calculate and plot dust attenuation curves for each galaxy. 

% Our code will be highly useful for theorists with galaxy evolution models that either do not have a dust prescription or would like a more empirically motivated one and wish to compare their simulations to observable features like photometry. 

This code can be used to add empirically motivated dust attenuation to theoretical models. With a bit of tweaking, it can also be used to generate narrower priors on dust attenuation parameters $n$ and $\tau$ for SED codes. With better and more restrictive priors on attenuation, especially with limited photometry and/or spectroscopy (compared to the 3D-HST photometric catalog), SED fits should have an easier time managing the dust-age-metallicity degeneracy.

\section{Conclusion} \label{sec:summary}

By utilizing posterior samples of SED fits from the state-of-the-art code \prosp of the rich 3D-HST data set, we have been able to create sophisticated population Bayesian models of dust attenuation as a function of stellar mass, SFR, metallicity, redshift, and axis ratio (proxy to inclination). 

We derive a qualitatively and quantitatively different relationship when taking the full galaxy posteriors and the \prosp priors into account (rather than simply adopting the maximum likelihood values), as shown in Figure \ref{fig:comp}. This shows that modeling highly correlated and prior-sensitive parameters is best done within a hierarchical framework. Our population model is also able to explain more of the scatter in the data than the simple Bayesian model.

% \prosp's dust attenuation model includes two components: diffuse dust, affecting the entire galaxy, and birth cloud dust, affecting only the most recently formed stars. As \prosp puts relatively strong constraints on the normalization of birth cloud dust based on that of diffuse dust, we have focused on the latter. 

We show the result of our model for diffuse dust optical depth as a function of birth cloud optical depth in Figure \ref{fig:dust1}. The relation is quasi-linear but considerably different from a one-to-one relation. The birth cloud dust optical depth increases faster than diffuse dust optical depth, which is in agreement with the picture painted in Figure 19 of \cite{Reddy2015} in conjunction with the connection between higher optical depths and higher SFRs \citep[e.g.,][]{Buat2009,Reddy2010,Reddy2015}.
% likely reflecting its direct dependence on current star formation processes.

% We also calculated ``posterior samples'' for an effective, one-component dust model and created models for those. An effective dust attenuation law is useful when detailed knowledge about individual stellar populations comprising a galaxy is not available. 

% Figure \ref{fig:diffeff} shows the direct comparison between diffuse and effective (single-component) dust attenuation curves. In general, the effective curve is stronger (obvious) and steeper than the diffuse curve. The extra steepness stems from the fact that young, massive stars that dominate the UV spectrum suffer extinction from both birth cloud and diffuse dust.

% The parameterization used for the diffuse and effective dust attenuation curve is given by Equation \ref{eq:diffdust}. In particular, the parameters that are allowed to vary are the normalization $\tau$ and attenuation slope $n$. In our population approach, we model $\tau$ and $n$ jointly as two separate general interpolations with a bivariate Gaussian intrinsic scatter (Equations \ref{eq:npoly} and \ref{eq:tau2poly}).

% Because $n$ depends most strongly on $\tau$, we compute both the bivariate models for $n$ and $\tau$ as well as univariate models for $n$ where $\tau$ is one of the independent variables. While the practical scenarios in which the univariate models can be applied are limited, they still yield highly interesting results that can be used to constrain theoretical dust models.

We then highlight some of the interesting results and implications of our bivariate (dependent variables optical depth $\tau$ and slope $n$) and univariate (variable $n$) models. We mainly focus on the effective dust attenuation models, but the diffuse dust generally behaves quite similarly. Here are the important points gleaned from our studies.

\begin{enumerate}
    \item The effective dust attenuation slope flattens with increasing optical depth, in agreement with the literature \citep[e.g.,][and references therein]{SalimNarayanan2020}. However, the expected relation between slope and $A_V$ is flatter than in previous studies, which is possibly due to \prosp's novel empirical, non-parametric star formation history, which led to lower current SFRs \citep{Leja2019}. As a result, the birth cloud dust contribution is less pronounced, signifying shallower attenuation curves. 
    % At high $A_V$, geometry likely plays a more important role in deciding the slope, thus reducing the differences from SED fitting choices.
    % \item Our univariate model for $n_{\rm eff}$ produces a very similar expectation curve as direct \prosp posteriors (no model application), but features a substantially reduced galaxy-to-galaxy scatter, highlighting the usefulness of our population model.
    \item The optical depth $\tau$ is most strongly correlated with SFR (positive). However, for $\log~{\rm SFR}\lesssim 0$, $\tau$ is nearly constant, even over a wide range of stellar masses.
    % , which is unsurprising given the fundamental connection between dust formation/destruction and star formation processes. This is likely due to a steady equilibrium state between formation and destruction of diffuse dust.
    \item Edge-on galaxies tend to have higher $\tau$ than face-on galaxies, but only for $\log~M_*\gtrsim 10$. We certainly expect such a relation, as for an edge-on galaxy, the line-of-sight penetrates through a significantly larger dust column density. However, low-mass galaxies lack disk-like morphologies. In other words, the relation between $b/a$ and viewing inclination is obfuscated.
    \item Dust attenuation curves evolve strongly with redshift at $z\gtrsim 1$ for low-mass, low-SFR galaxies, with stronger but flatter curves at increasing redshift. However, for $z>2$ we would need to push down to lower mass limits in the data to make a firm statement.
    % This may be caused by increasing pristine gas and dust reservoirs at higher redshift as the galaxies get farther below the SFMS.
    % This may be caused by the fact that such galaxies closer to the local universe lack high redshift analogs. The dust attenuation curves of such galaxies would be different from (common) low-mass, low-SFR galaxies at high redshift.
    % as they have often completed most or all of their possible star formation due to processes like excessive feedback or environmental effects (including ram pressure stripping)
    \item The relation between $n$ and $\log~M_*$ is multifaceted (the slope first flattens and then steepens with mass) and difficult to explain. 
    % We suggest an explanation involving differential obscuration of massive and lower-mass stars in addition to changing clumpiness of dust.
    % As shown earlier, attenuation slope is most strongly connected with optical depth, with shallower slopes at high $A_V$. The slope does depend on other parameters, including stellar mass, in highly complex ways. 
\end{enumerate}

Dust attenuation is an important aspect of integrated galaxy SEDs that must be properly accounted for in order to accurately measure other physical properties like SFR. However, its effects on the SED are similar to those of metallicity and age. With often limited photometry, SED fitting codes struggle to break these degeneracies. Meanwhile, ensemble dust attenuation studies are difficult to apply to individual galaxies due to the large variations found in the universe. Finally, there is no theoretical model of galaxy formation and evolution that treats dust from first principles all the way from the beginning to the current stage of the galaxy and predicts a reliable dust attenuation curve.

By creating more sophisticated models of dust attenuation based on a statistically rigorous framework that accounts for the large errors and often strong covariances between parameters of SED fits, we are advancing the knowledge of how dust attenuation curves vary from galaxy to galaxy based on their physical properties. Our results, accessible through the \texttt{Python} module \texttt{DustE} (\S \ref{sec:code}), can be used directly by modelers who wish to include the effects of dust when predicting observables without performing (full or approximate) radiative transfer. In addition, they can be used as stronger priors for individual SED fits in order to ensure more well-behaved posteriors, even when the likelihoods are poorly constrained by a lack of narrow- or medium-band multiwavelength photometry.

We have used the 3D-HST data set, with galaxies at $0.5<z<3.0$ over a wide variety of stellar masses, SFRs, and metallicities. In this paper, we have begun to explore the scientific implications of our population models regarding dust attenuation throughout the universe. With five dimensions and a flexible, essentially non-parametric functional form, we are able to register both expected and surprising trends. Our results can be used as tests of dust evolution models.

Further questions remain. First, a thorough theoretical analysis needs to be performed to explain some of the intriguing results we presented. Beyond the scope of this work, we restricted our analysis to the mass-complete sample, but employing treatments for incompleteness and pushing down to faint, lower-mass galaxies may lead to surprising changes in our picture of dust attenuation. Also, applying our hierarchical framework to a local set of galaxies, perhaps including a greater fraction of massive quiescent galaxies, could possibly yield different and equally interesting science results. In addition, simultaneously fitting dust emission using far-IR data \citep[e.g.,][]{Nagaraj2021b} may provide further constraints on or even different results for dust attenuation curves. 

\acknowledgments

This material is based upon work supported by the National Science Foundation Graduate Research Fellowship under Grant No. DGE1255832. Any opinion, findings, and conclusions or recommendations expressed in this material are those of the authors(s) and do not necessarily reflect the views of the National Science Foundation. The Flatiron Institute is supported by the Simons Foundation. We thank Rachel Somerville, Tjitske Starkenburg, and the CCA's Astronomical Data Group for helpful discussions.


\vspace{5mm}
\facilities{HST (WFC3), Spitzer (MIPS)}


\software{\texttt{pymc3} \citep{Pymc3}, AstroPy \citep{Astropy2013,Astropy2018}, SciPy \citep{Scipy2001,Scipy2020}, FSPS \citep{Conroy2009,Conroy2010SPSM}, \prosp \citep{Leja2017,Leja2019}, \texttt{dynesty} \citep{Speagle2020} }

\nocite{Somerville2015,Behroozi2013,Lilly2013,Han2014,daCunha2008,Fritz2007,ChevallardBeagle2016,Forbes2019}



\vspace{4cm}

\appendix

\section{Library of Terms} \label{sec:terms}

In this paper, we have defined many symbols with varying levels of abstraction. To help the reader keep track, we present a glossary that includes concrete examples of abstract terms in Table \ref{tab:glossary}.

\begin{table}[h!]
\resizebox{\hsize}{!}{
\hskip-2.0cm\begin{tabular}{@{}|l|l|l|l|@{}}
\toprule
Symbol & Meaning & Examples & \S \\ \midrule
$\{\mathbf{w}_k\}$ & Set of all true parameters of all individual galaxies & $\log~M_*$, SFR, AGN Fraction & \ref{subsec:probmath} \\ \midrule
$\{\mathbf{t}_k\}$ & Set of (true) parameters in our population model that are derived from other parameters in \prosp & $\log~M_*$, SFR & \ref{subsec:probmath} \\ \midrule
$\{\mathbf{u}_k\}$ & Set of (true) parameters in our population model that are fitted directly in \prosp & $\tau_2$, $n$ & \ref{subsec:probmath} \\ \midrule
$\{\mathbf{v}_k\}$ & Set of (true) parameters in \prosp not used in our population model that are used to compute $\{\mathbf{t}_k\}$ & SFH Parameters & \ref{subsec:probmath} \\ \midrule
$\{\mathbf{y}_k\}$ & Set of (true) parameters in \prosp completely unrelated to those we care about in the population model & AGN Fraction & \ref{subsec:probmath} \\ \midrule
$\{\mathbf{x}_k\}$ & Set of all data used in \prosp to fit individual galaxies & U-band photometry & \ref{subsec:probmath} \\ \midrule
$\boldsymbol{\theta}$ & Set of all hyperparameters dealing with the population of galaxies & $\sigma_1$ (see below), $M_*$ distribution parameters & \ref{subsec:probmath} \\ \midrule
$\boldsymbol{\alpha}$ & Set of all parameters defining \prosp prior distributions (fixed) & Min, max values for $n$ uniform distribution & \ref{subsec:probmath} \\ \midrule
$\tau_1$ & Birth cloud dust optical depth: normalization of Equation \ref{eq:tau1} & -- & \ref{subsec:daparam} \\ \midrule
$\tau_2$/$\tau_{\rm eff}$ & Diffuse/Effective dust optical depth: normalization of Equation \ref{eq:diffdust} & -- & \ref{subsec:daparam} \\ \midrule
$n$/$n_{\rm eff}$ & Diffuse/Effective dust slope parameter: see Equation \ref{eq:diffdust}. Higher values mean flatter curves & -- & \ref{subsec:daparam} \\ \midrule
S & Alternate slope parameter: $S=A_{1600}/A_V$. Higher values mean steeper curves & -- & \ref{subsec:AvS} \\ \midrule
$I_1$, $I_2$ & Linear interpolation functions used in population dust attenuation model & -- & \ref{subsec:moddetails} \\ \midrule
$\sigma_1$, $\sigma_2$, $\rho$ & Parameters of bivariate Gaussian distribution describing intrinsic scatter in the interpolation model & -- & \ref{subsec:moddetails} \\ \midrule
$\mathcal{L}$ & Statistical likelihood & -- & \ref{subsec:moddetails} \\ \bottomrule
\end{tabular}}
\caption{Glossary of mathematical symbols used in the paper. Intermediate quantities used to simplify equations and rarely mentioned terms are not included. For convenience, we also include the first section of appearance.}
\label{tab:glossary}
\end{table}

\section{\prosp Prior Sampling} \label{sec:prior}

In \S \ref{subsec:probmath}, we discussed that while the \prosp interim priors are defined as $p(\mathbf{u},\mathbf{v},\mathbf{y}|\boldsymbol{\alpha})$ (see Appendix \ref{sec:terms} for definitions), we would like to compute $p(\mathbf{u},\mathbf{t}|\boldsymbol{\alpha})$ instead considering $\mathbf{u}$ and $\mathbf{t}$ are the variables involved in our population model. To estimate $p(\mathbf{u},\mathbf{t}|\boldsymbol{\alpha})$, we sample directly from the \prosp priors without any data and empirically compute the probability distribution function (PDF) $p(\mathbf{u},\mathbf{t}|\boldsymbol{\alpha})$.

We use kernel density estimation (KDE) techniques to calculate $p(\mathbf{u},\mathbf{t}|\boldsymbol{\alpha})$ for the variables we are considering in the particular population model. Parameters with uniform priors are ignored as they would simply add constants in log space. For example, for the bivariate model of $(n_{\rm eff},\tau_{\rm eff})$ as a function of stellar mass, SFR, stellar metallicity, redshift, and axis ratio; we calculate the joint prior PDF for stellar mass, SFR, metallicity, and $\tau_{\rm eff}$, which have non-uniform priors. We use a Gaussian kernel with a bandwidth of $0.2*n-0.1$, where $n$ is the number of variables in the prior. This functional form was decided empirically based on optimization studies on prior sub-samples.

We have sampled from the \prosp priors ($N=500,000$) at six evenly-spaced redshifts from $z=0.5$ to $z=3.0$. We construct a PDF at each redshift using KDE and use a linear interpolation to calculate the effective joint prior PDF for any given galaxy sample given the redshift.

\begin{figure*}
    \centering
    \resizebox{\hsize}{!}{
    \includegraphics{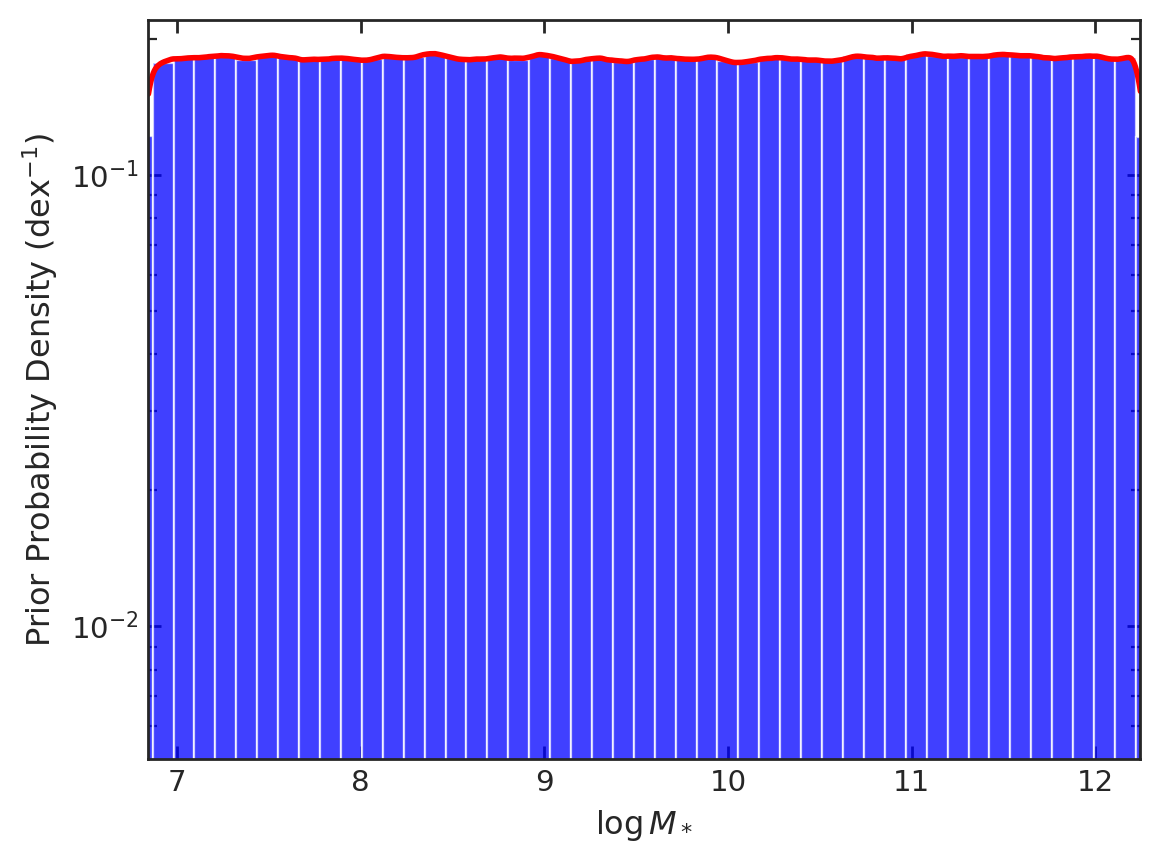}
    \includegraphics{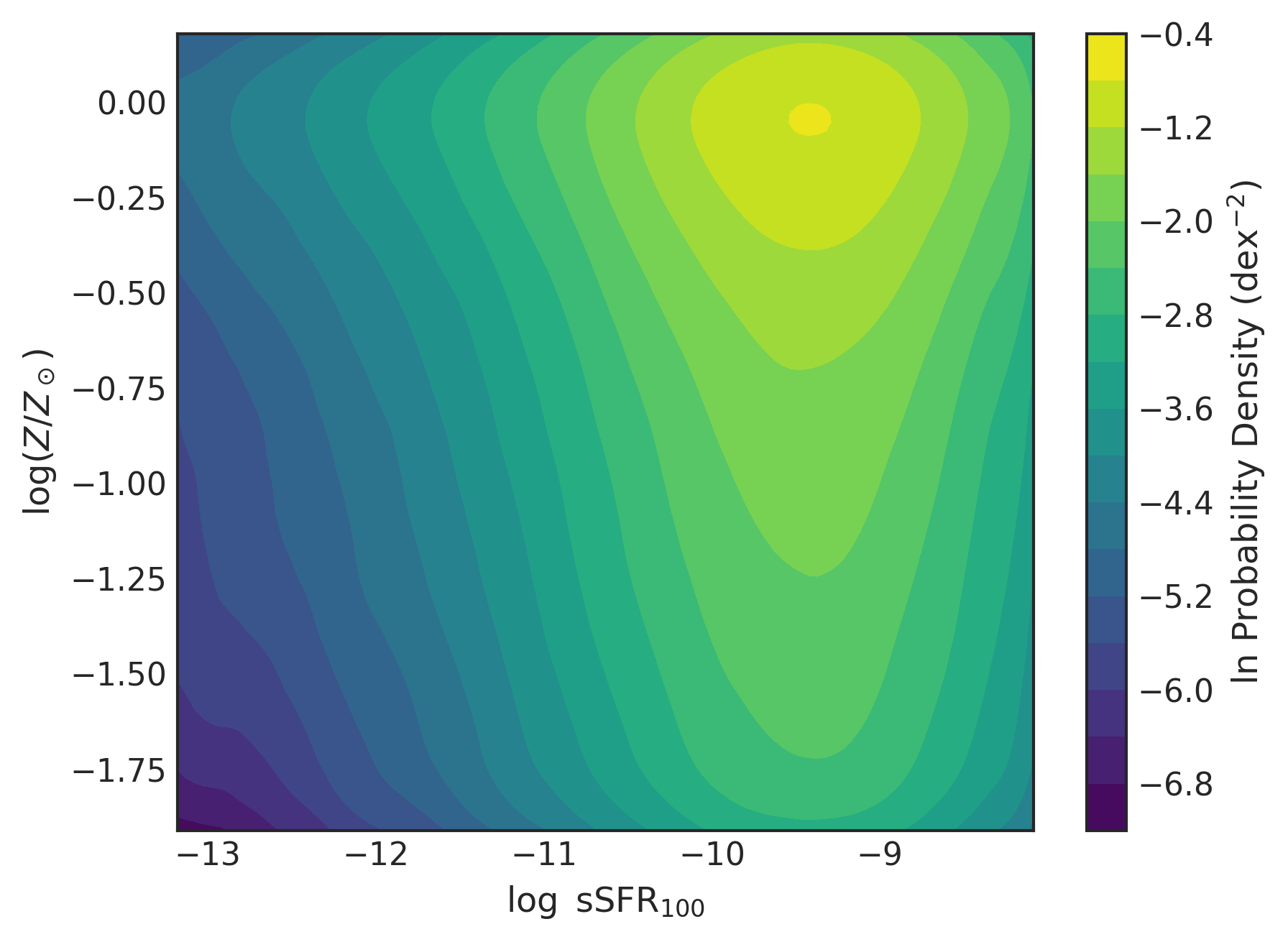}}
    \caption{Prior PDF with respect to stellar mass (as a single variable) at redshift $z=0.5$ (left panel) and with respect to sSFR and metallicity (as a double variable) at $z=3$ (right panel). In the left panel, we see that the PDF (red line) traces the prior sample histogram (blue) very well.}
    \label{fig:priors}
\end{figure*}

In Figure \ref{fig:priors}, we show the prior PDF for stellar mass (single variable) at redshift $z=0.5$ (left panel) and the PDF for specific SFR (sSFR) and metallicity (two variables) at redshift $z=3.0$ in the right panel. The stellar mass PDF is nearly uniform, which is what we expect since stellar mass is closely related to the total mass formed over a galaxy's history, which has a uniform prior distribution in \prosp. In the right panel, we see that according to the \prosp priors, at $z=3$ higher sSFRs and metallicities are more probable than lower ones. 

\section{Validation of the Population Methodology} \label{sec:simres}

To ensure that our population model works properly, we simulated data and tested how well the model could reproduce the truth. We selected anywhere from one to four independent variables (from the list of stellar mass, sSFR, metallicity, redshift, and axis ratio, though only the number of variables is important). The independent variable values were chosen from a uniform random distribution. 

% so that the median value of each variable was $0$. For example, a log stellar mass distribution $[8,12]$ with median $9.5$ is represented as $[-1.5,2.5]$. We chose to center the polynomials around the medians of the independent variables as this lead to much more reasonable and stably computed coefficients.

The true values of the dependent variables $n$ and $\tau$ were formulated with N-D degree-2 polynomials of the independent variables such that the values were always within the bounds introduced in \S \ref{subsec:moddetails}. Scatter was added to the polynomials based on the true values of $\log \sigma_1$, $\log \sigma_2$, and $\rho$ as indicated by the equations in \S \ref{subsec:config}.

% The values of the dependent variables $n$ and $\tau_2$ were chosen in the following manner. Based on the polynomial degree and number of independent variables, a coarse grid with the same number of points as coefficients was created that straddled the entire space of independent variables. The values of $n$ and $\tau_2$ were chosen randomly from their prior distributions in \prosp (see \S \ref{subsec:simres}). For any particular random sample, a specific polynomial is specified, thus creating a truth. Scatter was added to the polynomials based on true values of $\sigma_1$, $\sigma_2$, and $\rho$ as indicated by the equations in \S \ref{subsec:config}.

To create ``posterior samples'' for the simulated objects, we added a Gaussian random variable with mean of zero and specified $\sigma$ to shift the true values of both independent and dependent variables to an observed mean. Then, we added another Gaussian random variable with mean of zero and the same $\sigma$ to create samples around the observed mean. The ratio of $\sigma$ values for the different variables was set to mimic the \prosp data (based on the average standard deviations of the true samples), but the absolute value was determined on the command line in order to test how well our algorithm works for both clean and noisy data.

We then modeled the simulated dust attenuation parameters (\S \ref{subsec:daparam}) using the equations and techniques described in \S \ref{subsec:config}, using the interpolation model with the same set of independent variables.

\begin{figure*}
    \centering
    \resizebox{\hsize}{!}{
    \includegraphics{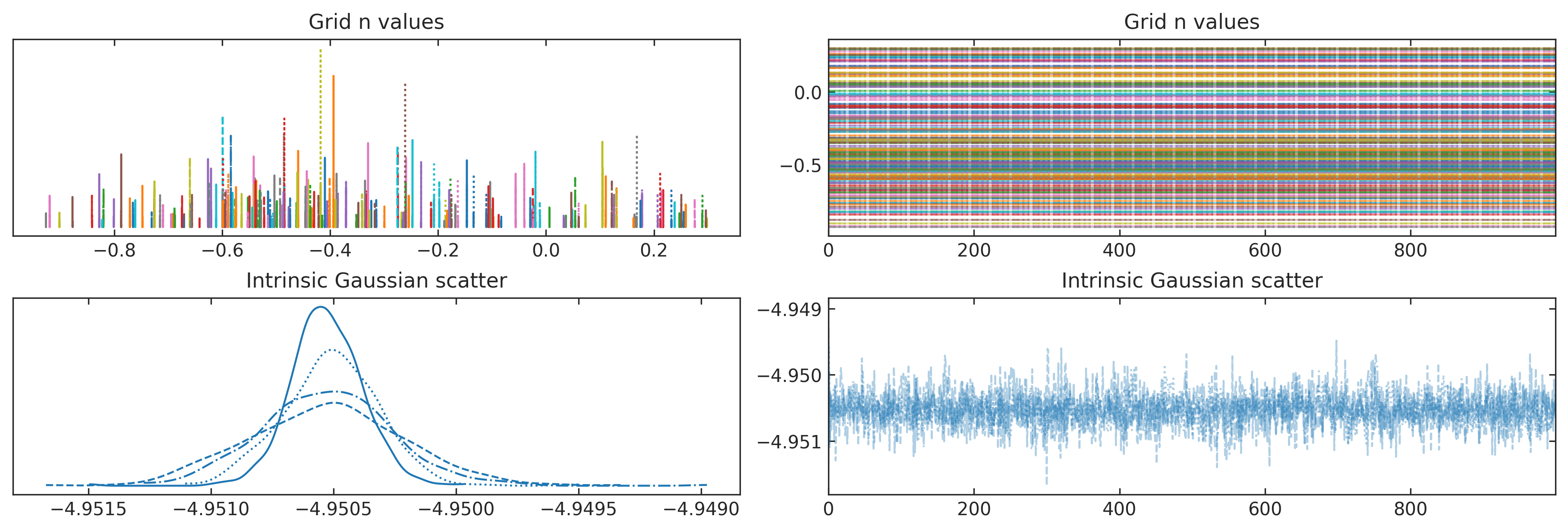}}
    \caption{Kernel density estimation (KDE) plots (left panels) and trace plots (right panels) for simulation of dust attenuation as a degree-2 polynomial (truth) in stellar mass and sSFR. The KDE plots show the posterior probability of the given parameter as a function of that value. Different chains are shown with different line styles. The top parameter, "Grid n values," is in fact a conglomeration of several variables (the value of $n$ at different parts of the grid), each one with a different color. The convergence for each parameter within each chain and between chains is made clear through well-defined peaks with Gaussian tails (especially visible for the scatter parameter) and the overlap of same-color curves with different line styles. The trace plots show the value of each variable over the duration of each chain. The variables vary within a fraction of a percent over the entire duration of the chains, showing a very stable solution.}
    \label{fig:trace}
\end{figure*}

In essentially all cases, our population model is able to retrieve the true parameter values within reasonable error bounds. Here, we show the results for a dust attenuation model of degree 2 (truth) as a function of stellar mass and sSFR using 6000 simulated galaxies. In Figure \ref{fig:trace}, we show the posterior probability (left panels) of two parameters in the model (values of $n$ on the grid and the fixed intrinsic Gaussian scatter for $n$) as well as their values throughout the NUTS Markov chains (right panels). Given the uniformity in each parameter within each chain (clear posterior probability peaks with Gaussian tails and fraction-of-a-percent change in the parameters over the chains' duration) and between chains (overlap between differently-styled lines of the same color), we can see that the code has converged successfully.

\begin{figure*}
    \centering
    \resizebox{\hsize}{!}{
    \includegraphics{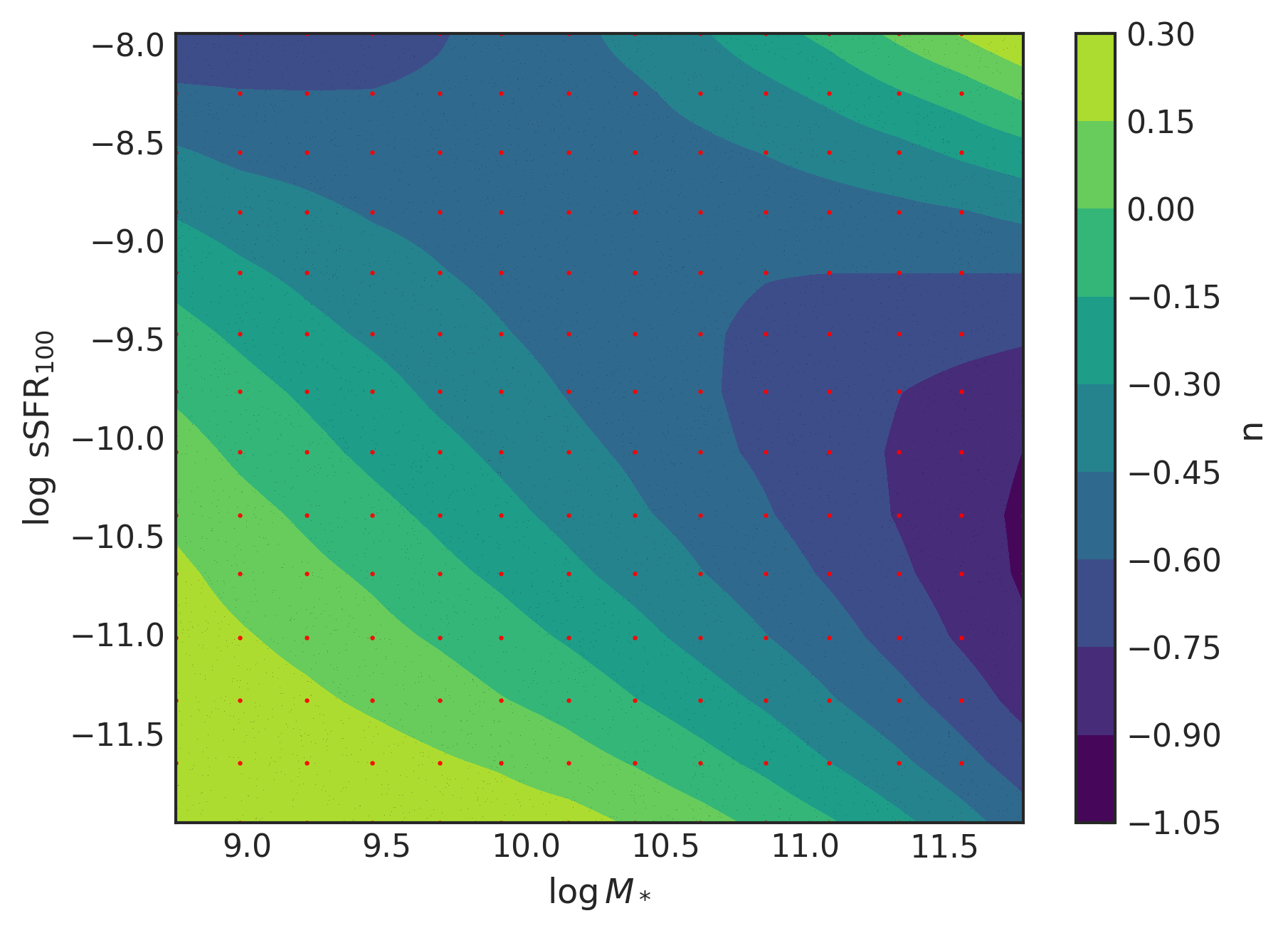}
    \includegraphics{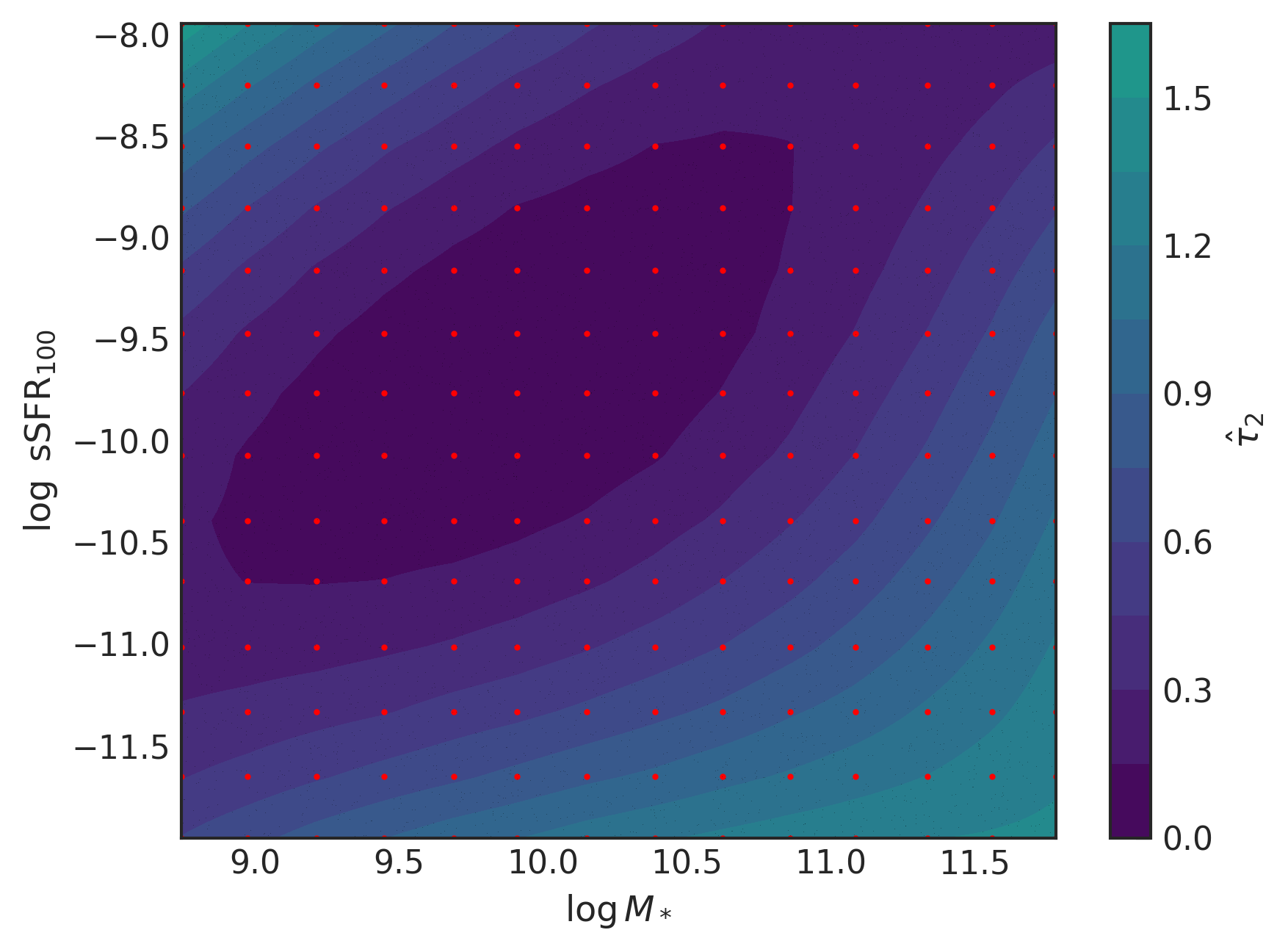}}
    \resizebox{\hsize}{!}{
    \includegraphics{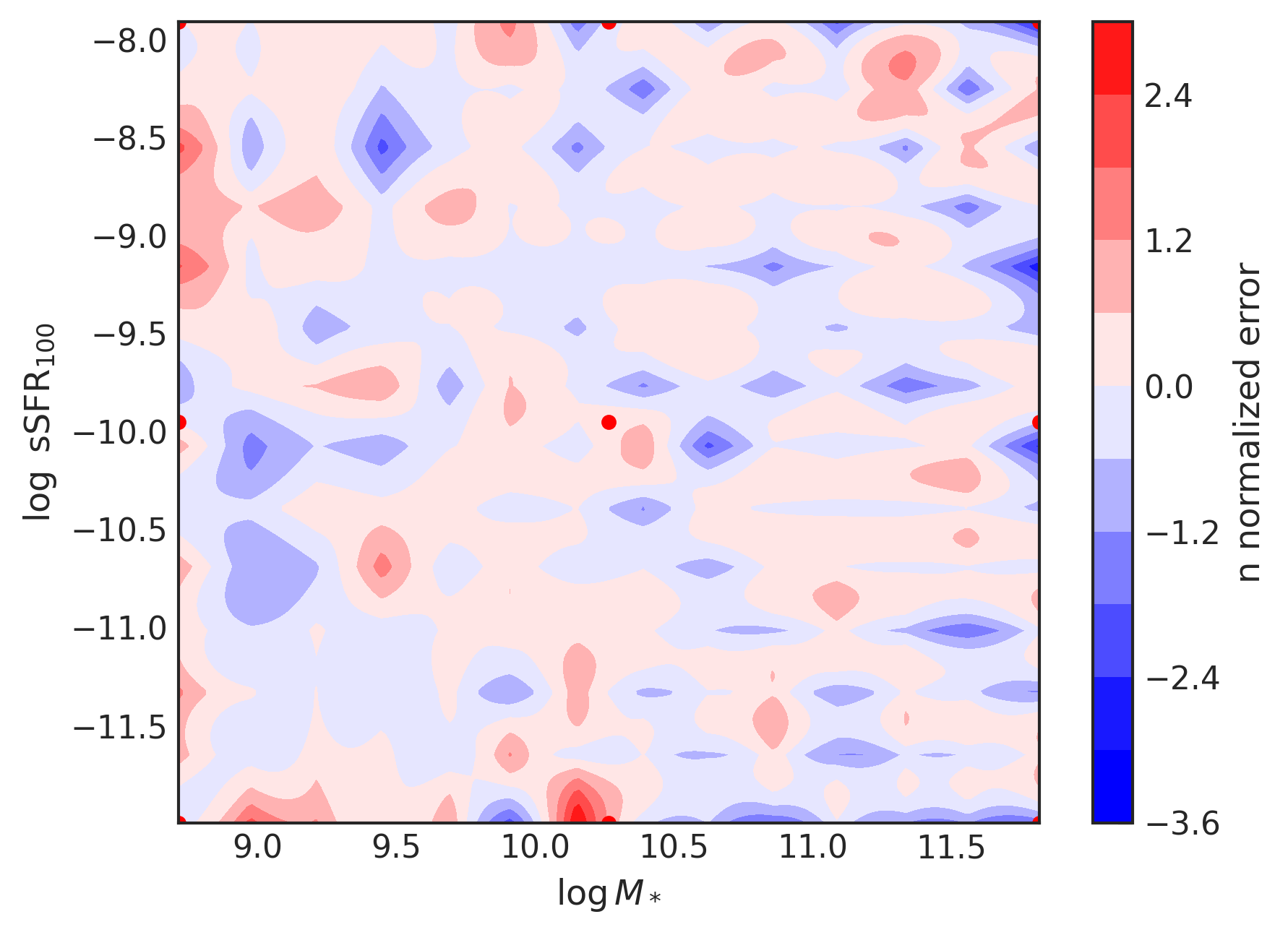}
    \includegraphics{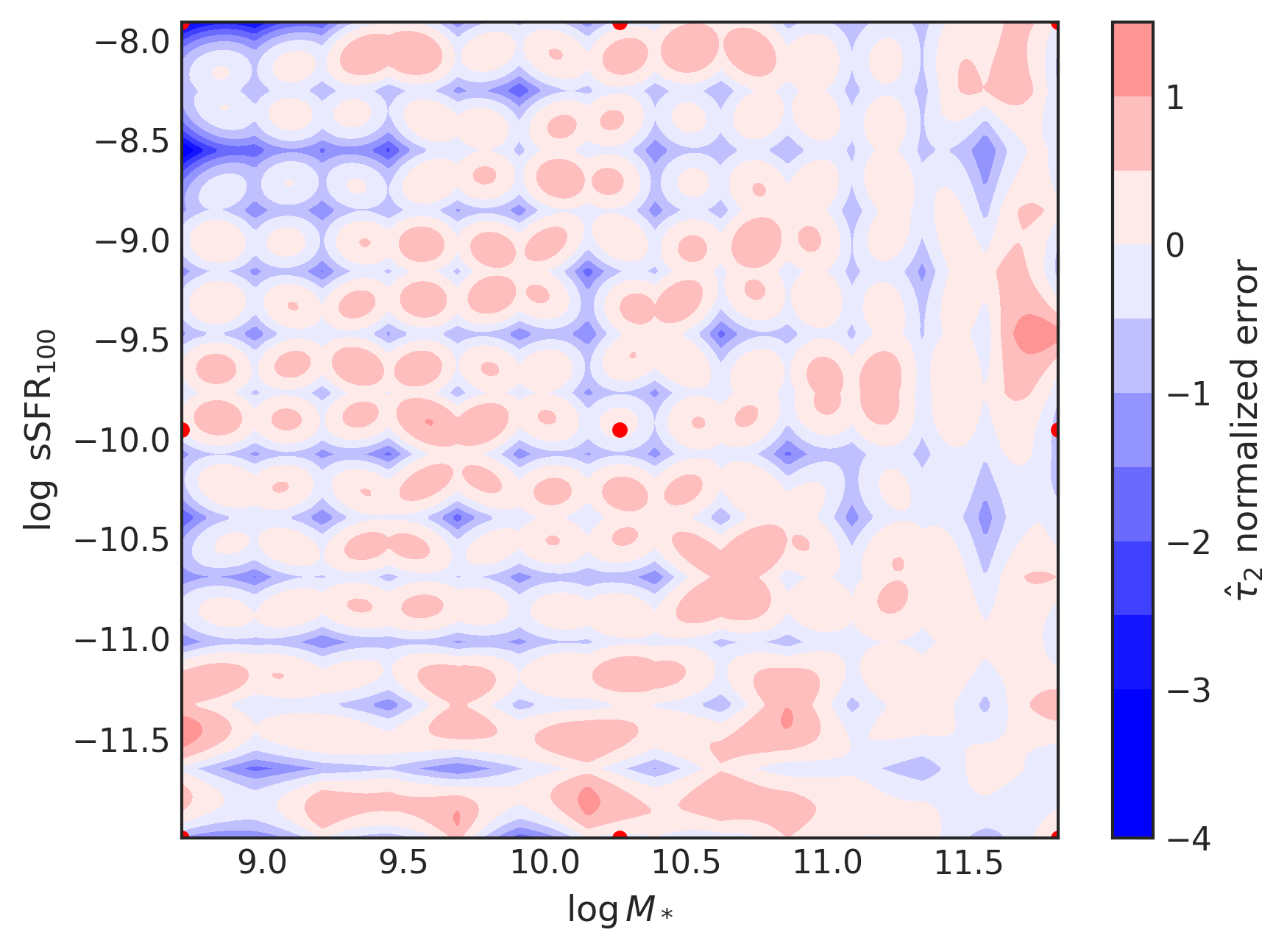}}
    \resizebox{\hsize}{!}{
    \includegraphics{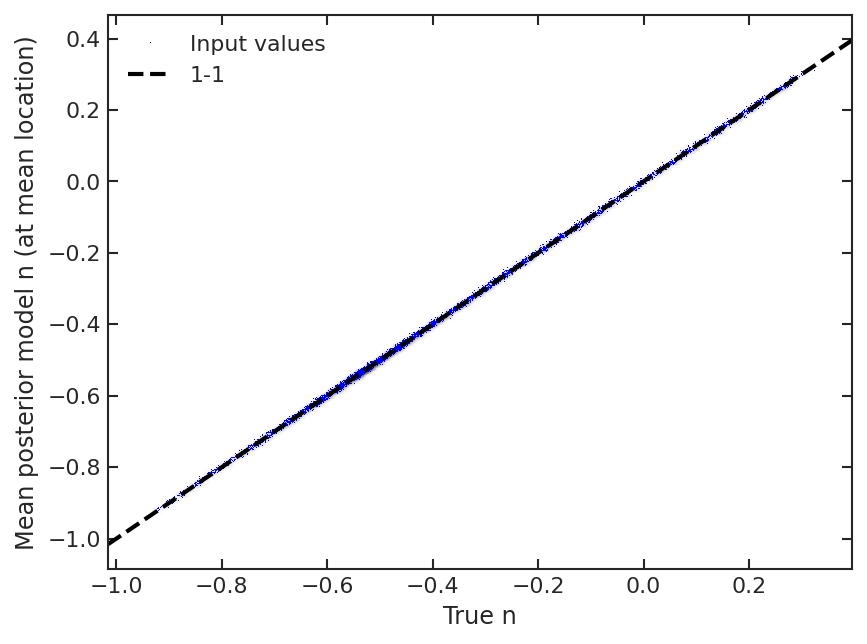}
    \includegraphics{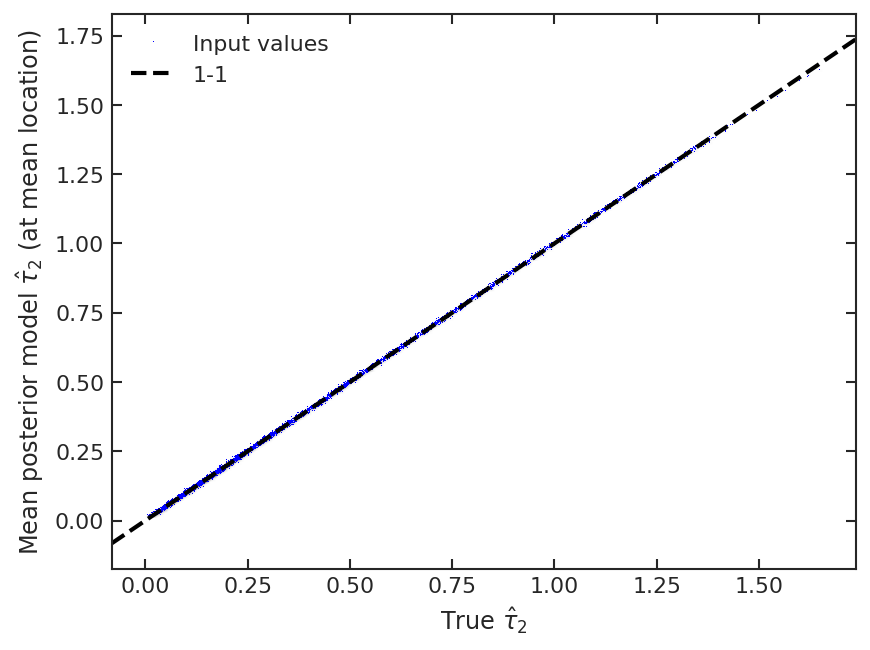}}
    \caption{Test of our population modeling capabilities. For the true dust attenuation, we set both $n$ and $\tau_2$ as degree-2 polynomials of log stellar mass and log sSFR and create 6000 ``galaxies.'' We simulate with our interpolation model. The top two panels show the mean model for $n$ (left) and $\tau_2$ (right) as a function of $\log~M_*$ and $\log {\rm sSFR}$. The middle two panels show the normalized residual maps (difference between model and truth divided by overall intrinsic scatter) in the same parameter space. The red dots in this case represent the grid on which the polynomial model was calculated. The lack of visible systematics suggests that the model is successful. In this particular model, we do slightly overestimate the intrinsic scatter (simulated with $\log \sigma_1=\log \sigma_2=-5.30$ but recovered $\log \sigma_1=-4.95$ and $\log \sigma_2=-5.09$), but given how small the ``measurement errors'' are ($\sim 5\%$ of the \prosp errors), this is not a cause for concern. The bottom two panels show the comparison between true $n$ (left) and $\tau_2$ (right) parameters and model parameters. We see excellent agreement between the truth and the model, especially for $n$.}
    \label{fig:simres}
\end{figure*}

Next, we show the actual fitted models and comparison between true and model parameter values in Figure \ref{fig:simres}. The model recovers the truth very well, as evidenced by the middle and bottom panels. The bottom panels show the true (``posteriors'') vs measured $n$ and $\tau$ values, which are consistent with a one-to-one relationship. The middle panels show the normalized residuals (difference between modeled and true divided by the quadratic sum of intrinsic scatter in the truth and model). 

There are no visible systematics. We do slightly overestimate the intrinsic scatter in the relationship ($\log \sigma_1=\log \sigma_2=-5.30$ truth vs recovered values of $\log \sigma_1=-4.95$ and $\log \sigma_2=-5.09$). This $\sim 30\%$ overestimate in intrinsic scatter arises because the simulation is unable to perfectly distinguish between measurement errors and intrinsic scatter when the errors (taken to be of the same order of magnitude as the intrinsic scatter) are only $5\%$ of the \prosp measurement errors. We find for measurement errors on the order of those in \prosp (not plotted in this paper), as long as the intrinsic scatter is over $\sim 20\%$ of the typical \prosp errors, we recover even the scatter with no bias.

% We do find that modeling with a different set of independent variables and/or polynomial degree than was used to create the truth leads to sub-optimal results (in trying to reproduce the posterior samples of the dependent variables). In particular, modeling with fewer independent variables and/or a lower polynomial degree results in bad fits whereas modeling with more independent variables and/or a higher polynomial degree tends to reproduce the posterior samples somewhat.

\section{Marginalization} \label{sec:marg}

% As our models have five independent variables, we need a process to compute lower-dimensional cross-sections of the results for easier visualization. In this section, we derive the marginalization process. From the equations, we find that accounting for underlying priors assumed as well as the distribution of the data is important to guarantee statistical rigor.

% Suppose the set of all independent variables is $\theta$. We let the subset of variables over which we want to marginalize be $\theta_m$, and the remaining parameters $\theta_x$. We consider the dependent variable $n(\theta)$. 

% % When we remove boldface, we are referring to particular values of a given set of variables. 

% The quantity we would like to calculate is the expected value of the function $n(\theta_x)$ given $\theta_m$, or $E\left[n_{\theta_m}(\theta_x) \right]$. Our model function also depends on the 3D-HST data $\mathbf{X}$ via the \prosp  In addition, we must take into account the \prosp prior parameters $\alpha$ as conditioned on the 3D-HST data $\mathbf{X}$. This introduces a $p(\alpha | \mathbf{X}$) dependence in our answer.

% \begin{equation}
%     E\left[n_{\theta_m}(\theta_x) \right] = \int n_{\theta_m}(\theta_x) p(\theta_m,\alpha|\mathbf{X})d\theta d\alpha
% \end{equation}

% \subsection{Attempt at a Pedagogical Version}

At various points in this work, we have visualized our model in a lower-dimensional space than its native 5 dimensions of independent parameters. Here we explain the not-necessarily-straightforward choices we make to do so.

The hyperparameters of our model, $\boldsymbol{\theta}$, are mostly the values of the dependent variable(s) at control points used for interpolation. Given a particular choice of $\boldsymbol{\theta}$, our model predicts a distribution of some dependent variable(s) at every point in the five-dimensional space of independent variables. The independent variables are things like $M_*$, SFR, and so on, and the dependent variables are $n$, $n_\mathrm{eff}$, $n$ and $\tau$, or $n_\mathrm{eff}$ and $\tau_\mathrm{eff}$. When we visualize the model's predictions as a function of, say, just $M_*$ and SFR, we have to make a choice about what to do with the remaining dimensions, e.g. $z$, $b/a$, and $Z_*$. In analogy with visualizing a 3D cube of data in 2 dimensions, one could imagine 2 basic options. The first is to fix the values of these other parameters to create a 2D slice of the prediction. The second is to integrate over the other parameters to create a projection, possibly with some weight other than simply the ``volume'' of the parameter space, which may be somewhat arbitrary, e.g. when integrating over $M_*$, should one integrate over $M_*$ or $\log M_*$?

To avoid slices which would presumably be quite sensitive to the particular fixed values of the suppressed independent parameters, we have chosen to use a projection. In some sense, therefore, our task now is to choose a reasonable weight function and elucidate exactly how we approximate the corresponding integrals as sums. We also need to keep in mind that the model predicts a gaussian distribution at every point in the 5-dimensional independent parameter space, not just a single value.

Let's define the model's predicted PDF at such a point in parameter space to be $f(\delta | {\bf X}, \boldsymbol{\theta})$. The dust parameters are denoted $\delta$ (which will be one- or two-dimensional depending on the flavor of the model), the point in the 5-dimensional space of independent variables is ${\bf X}$, and the parameters of the population model are $\boldsymbol{\theta}$. At a fixed value of ${\bf X}$, there is no ambiguity about the mean value of $\delta$:
\begin{equation}
    E_{\boldsymbol{\theta}|\{\mathbf{x}_k\}}[\delta | {\bf X}] = \int \delta\ f(\delta | {\bf X}, \boldsymbol{\theta}) p(\boldsymbol{\theta} | \{\mathbf{x}_k\})  d\boldsymbol{\theta} d\delta.
\end{equation}
The integral over $\delta$ is necessary because $f$ is itself a PDF. The second factor, $p(\boldsymbol{\theta} | \{\mathbf{x}_k\})$ is the posterior distribution of the model's hyperparameters $\boldsymbol{\theta}$. Using the usual monte carlo approximation of this integral, we would estimate
\begin{equation}
\label{eq:marg5d}
    E_{\boldsymbol{\theta}|\{\mathbf{x}_k\}}[\delta | {\bf X}] \approx \int M^{-1} \sum_{m=1}^M \delta\ f(\delta | {\bf X}, \boldsymbol{\theta}^{(m)}) d\delta \approx M^{-1} \sum_{m=1}^M \delta_{\mathrm{avg},\boldsymbol{\theta}^{(m)}},
\end{equation}
where $m$ indexes the $M$ posterior samples of our population model, and the second approximation holds if the width of the distribution $f$ is insensitive to $\boldsymbol{\theta}$, which we find to be the case in practice. $\delta_\mathrm{avg}$ is a straightforward function of $\boldsymbol{\theta}$, i.e. the result of the N-dimensional interpolations, so this approximation is particularly convenient.

If we now split ${\bf X}$ into ${\bf X}_v$ and ${\bf X}_s$, which will be respectively visualized and suppressed, the quantity we should compute is of the form
\begin{equation}
    E_{{\bf X}_s,\boldsymbol{\theta}|\{\mathbf{x}_k\}}[\delta | {\bf X}_v] = \int \delta\ f(\delta | {\bf X}, \boldsymbol{\theta}) p(\boldsymbol{\theta} | \{\mathbf{x}_k\}) W({\bf X}_s | \boldsymbol{\theta}, \{\mathbf{x}_k\})  d\boldsymbol{\theta} d\delta d{\bf X}_s.
\end{equation}
A reasonable choice for the weight function $W({\bf X}_s | \boldsymbol{\theta}, \{x_i\})$, which in the general case may depend on $\boldsymbol{\theta}$ and $\{x_i\}$, is the density of data points. In principle this should have the advantage of not incorporating regions of parameter space with large uncertainty, i.e. regions not constrained by data, into the visualized averages. However, the true density of galaxies at any point ${\bf X}$ is nontrivial to determine \citep[see][]{ForemanMackey2014HierBay} given the uncertainties in the data. These uncertainties are encapsulated in the posterior samples we have from \prosp for each galaxy, which are taken to be drawn from the distribution $p(\{\mathbf{t}_k\},\{\mathbf{u}_k\} | \{\mathbf{x}_k\},\boldsymbol{\alpha})$ (related to Equation \ref{eq:postdist}), where again $\boldsymbol{\alpha}$ represents the parameters controlling the interim prior used by \prosp. Note that ${\bf X}$ is a subset of the parameters sets $\mathbf{t}_k$ and $\mathbf{u}_k$, so $p({\bf X}_s | \{\mathbf{x}_k\},\boldsymbol{\alpha})$ is a marginal distribution of $p(\{\mathbf{t}_k\},\{\mathbf{u}_k\} | \{\mathbf{x}_k\},\boldsymbol{\alpha})$. However, just as in the main text, we are interested in this distribution with the effects of the interim prior $\boldsymbol{\alpha}$ removed. 

To approximate this integral, we employ similar methodology to our discussion in Section \ref{sec:methods}. In particular, we rely on sums over the \prosp samples to perform the integral over ${\bf X}_s$, sums over the posterior samples of the population model to integrate over $\boldsymbol{\theta}$, and given the relative lack of posterior uncertainty in the parameters controlling the width of the normal distributions $f(\delta | {\bf X},\boldsymbol{\theta})$, the integral over $\delta$ is approximated by neglecting this width as in the second approximation in Equation \eqref{eq:marg5d}. We therefore approximate this average as
\begin{equation}
    E_{{\bf X}_s,\boldsymbol{\theta}|\{\mathbf{x}_k\}}[\delta | {\bf X}_v] \approx M^{-1}Q^{-1} \sum_{m=1}^M \sum_{q=1}^Q \delta_{\mathrm{avg},\boldsymbol{\theta}^{(m)}} .
\end{equation}
Here again $m$ indexes samples from the population model. The $Q$ points, meanwhile, are drawn from the sample of $N$ galaxies times the (up to) 500 \prosp samples available per galaxy, with a weight equal to $1/p(\mathbf{t}_q,\mathbf{u}_q | \boldsymbol{\alpha})$, i.e. the inverse of the \prosp priors at the given sample point. In the language of slices and projections, this is a projection constructed by averaging many slices through the parameter space using a fine grid of ${\bf X}_v$ values, where the fixed values of the suppressed dimensions ${\bf X}_s$ are drawn from the population of \prosp posterior points with the \prosp priors removed.
% \begin{equation}
%     E\left[n_{\boldsymbol{\theta}_m}(\boldsymbol{\theta}_x) \right] = \int n_{\boldsymbol{\theta}_m}(\boldsymbol{\theta}_x) \frac{p(X|\boldsymbol{\theta}_m, \boldsymbol{\alpha})p(\boldsymbol{\theta}_m,\boldsymbol{\alpha})}{p(X)} d\boldsymbol{\theta} d\boldsymbol{\alpha}.
% \end{equation}

% \begin{equation}
%     E\left[n_{\boldsymbol{\theta}_m}(\boldsymbol{\theta}_x) \right] = \int n_{\boldsymbol{\theta}_m}(\boldsymbol{\theta}_x) \frac{p(X|\boldsymbol{\theta}_m, \boldsymbol{\alpha})p(\boldsymbol{\theta}_m,\boldsymbol{\alpha})}{p(X)}d\boldsymbol{\theta} d\boldsymbol{\alpha}.
% \end{equation}


\bibliography{sample63}{}
\bibliographystyle{aasjournal_mod}



\end{document}